\documentclass[
aip,
jcp,
amsmath,
amssymb,
preprint,
onecolumn,
superscriptaddress,
]{revtex4-1}

\usepackage{graphicx}
\usepackage{dcolumn}
\usepackage{bm}

\usepackage[utf8]{inputenc}
\usepackage[T1]{fontenc}
\usepackage{mathptmx}
\usepackage{etoolbox}

\makeatletter
\def\@email#1#2{%
 \endgroup
 \patchcmd{\titleblock@produce}
  {\frontmatter@RRAPformat}
  {\frontmatter@RRAPformat{\produce@RRAP{*#1\href{mailto:#2}{#2}}}\frontmatter@RRAPformat}
  {}{}
}%
\makeatother

\usepackage[english]{babel}
\usepackage[version=3]{mhchem}
\graphicspath{{IMG/}}
\usepackage{braket}
\usepackage{todonotes}
\usepackage[nohyperlinks,nolist]{acronym}
\usepackage{mathtools}
\usepackage{hyperref}
\usepackage[inline]{enumitem}
\usepackage{subfigure}
\usepackage{multirow}

\definecolor{bg}{rgb}{0.95,0.95,0.95}
\usepackage[frozencache=true,cachedir=minted-cache,newfloat]{minted}
\setminted{
  fontsize=\footnotesize,
  xleftmargin=\parindent,
  xrightmargin=\parindent,
  frame=lines,
  bgcolor=bg 
}

\newcommand{\vampyr}{\texttt{VAMPyR}}
\newcommand{\pylibxc}{\texttt{Pylibxc}}
\newcommand{\mrcpp}{\texttt{MRCPP}}
\newcommand{\mrchem}{\texttt{MRChem}}
\newcommand{\mrchemsoft}{\texttt{MRChemSoft}}
\newcommand{\veloxchem}{\texttt{VeloxChem}}
\newcommand{\madness}{\texttt{M-A-D-N-E-S-S}}
\newcommand{\numpy}{\texttt{NumPy}}
\newcommand{\scipy}{\texttt{SciPy}}
\newcommand{\matplotlib}{\texttt{Matplotlib}}
\newcommand{\functiontree}{\texttt{FunctionTree}}
\newcommand{\pybind}{\texttt{Pybind11}}

\begin{acronym}
\acro{MW}{MultiWavelet}
\acro{AO}{Atomic Orbital}
\acro{MO}{Molecular Orbital}
\acro{XC}{Exchange--Correlation}
\acro{LDA}{Local Density Approximation}
\acro{GGA}{Generalized Gradient Approximation}
\acro{GTO}{Gaussian-Type Orbital}
\acro{LCAO}{Linear Combination of Atomic Orbital}
\acro{MRA}{Multi--Resolution Analysis}
\acro{BSH}{Bound--State Helmholtz}
\acro{PCM}{Polarizable Continuum Model}
\acro{SCF}{Self--Consistent Field}
\acro{HF}{Hartree--Fock}
\acro{KS-DFT}{Kohn--Sham Density Functional Theory}
\acro{GIL}{global interpreter lock}
\acro{PES}{Potential Energy Surface}
\acro{NS}{non-standard}
\acro{KS}{Kohn--Sham}
\acro{SCRF}{self-consistent reaction field}
\acro{GPE}{generalized Poisson equation}
\end{acronym}

\makeatletter
\newcommand{\tutorial}[2]{\def\@currentlabel{#1}\label{#2}}
\makeatother

\tutorial{T1}{tut:mra}
\tutorial{T2}{tut:functions}
\tutorial{T3}{tut:image-comp}
\tutorial{T4}{tut:arithmetics}
\tutorial{T5}{tut:derivatives}
\tutorial{T6}{tut:convolutions}
\tutorial{T7}{tut:hydrogen}
\tutorial{T8}{tut:helium}
\tutorial{T9}{tut:beryllium}
\tutorial{T10}{tut:water}
\tutorial{T11}{tut:h2-dis}

\makeatletter
\renewcommand{\@cite}[1]{
{$\!\! ^{(#1)}$}}
\makeatother

\newcommand{\etal}{{\emph{et al.}\ }}

\newcommand{\orbital}{\ensuremath{\varphi}}

\newcommand{\fockMat}{\ensuremath{F}}
\newcommand{\fockOper}{\ensuremath{\hat{F}}}

\newcommand{\scaling}{\varphi}
\newcommand{\wavelet}{\psi}

\begin{document}

\title{\vampyr{} -- A High-Level Python Library for Mathematical Operations in a Multiwavelets Representation}

\author{Magnar Bj{\o}rgve*}
\affiliation{Hylleraas center, Department of Chemistry, UiT The Arctic University of Norway, PO Box 6050 Langnes, N-9037 Troms\o, Norway.}

\author{Christian Tantardini}
\affiliation{Hylleraas center, Department of Chemistry, UiT The Arctic University of Norway, PO Box 6050 Langnes, N-9037 Troms\o, Norway.}
\affiliation{Department of Materials Science and NanoEngineering, Rice University, Houston, Texas 77005, United States of America}

\author{Stig Rune Jensen}
\affiliation{Hylleraas center, Department of Chemistry, UiT The Arctic University of Norway, PO Box 6050 Langnes, N-9037 Troms\o, Norway.}

\author{Gabriel A. Gerez S.}
\affiliation{Hylleraas center, Department of Chemistry, UiT The Arctic University of Norway, PO Box 6050 Langnes, N-9037 Troms\o, Norway.}

\author{Peter Wind}
\affiliation{Hylleraas center, Department of Chemistry, UiT The Arctic University of Norway, PO Box 6050 Langnes, N-9037 Troms\o, Norway.}

\author{Roberto Di Remigio Eik\aa s}
\affiliation{Algorithmiq Ltd, Kanavakatu 3C, FI-00160 Helsinki, Finland}
\affiliation{Hylleraas center, Department of Chemistry, UiT The Arctic University of Norway, PO Box 6050 Langnes, N-9037 Troms\o, Norway.}

\author{Evgueni Dinvay}
\affiliation{Hylleraas center, Department of Chemistry, UiT The Arctic University of Norway, PO Box 6050 Langnes, N-9037 Troms\o, Norway.}

\author{Luca Frediani*}
\affiliation{Hylleraas center, Department of Chemistry, UiT The Arctic University of Norway, PO Box 6050 Langnes, N-9037 Troms\o, Norway.}

\email{magnbjor@gmail.com, luca.frediani@uit.no}

\date{\today}

\begin{abstract}
Wavelets and Multiwavelets have lately been adopted in Quantum Chemistry to overcome challenges presented by the two main families of basis sets: Gaussian atomic orbitals and plane waves. In addition to their numerical advantages (high precision, locality, fast algorithms for operator application, linear scaling with respect to system size, to mention a few) they provide a framework which narrows the gap between the theoretical formalism of the fundamental equations and the practical implementation in a working code. This realization led us to the development of the Python library called \vampyr{}, (Very Accurate Multiresolution Python Routines). \vampyr{} encodes the binding to a C++ library for Multiwavelet calculations (algebra, integral and differential operator application) and exposes the required functionality to write simple Python code to solve among others, the Hartree--Fock equations, the generalized Poisson Equation, the Dirac equation and the time-dependent Schrödinger equation up to any predefined precision. In this contribution we will outline the main features of Multiresolution Analysis using multiwavelets and we will describe the design of the code. A few illustrative examples will show the code capabilities and its interoperability with other software platforms.
\end{abstract}

\maketitle


\section{Introduction}
\label{sec:intro}

Wavelets and Multiwavelets have emerged in the last few decades as a versatile tool for computational science. Their strength derives from the combination of frequency separation with locality and a robust mathematical framework to gauge the precision of a calculation. In particular, Multiwavelets have been recently employed within the field of Quantum Chemistry to overcome some of the known drawbacks of traditional \ac{AO}-based calculations.\cite{Harrison2004-jz,Yanai2004-mc,Yanai2004-tw,Yanai2005-ve,Brakestad2021-fh,Jensen2014-lg,Jensen2016-vb,Jensen2017-ko,Pitteloud2023-lr} The code which pioneered this approach is \madness{},\cite{Harrison2016-lo} followed by our own code \mrchem{}.\cite{wind2022mrchem} To date, they are the only two codes available for quantum chemistry calculations using multiwavelets.

As the development of \mrchem{} unfolded, we found advantageous to separate the mathematical code dealing with the \ac{MW} formalism in a separate library, called \mrcpp{} (Multiresolution Computation Program Package) \cite{mrcpp_github}. \mrcpp{} is a high-performance C++ library, which provides the tools of \ac{MRA} with Multiwavelets. It implements low-scaling algorithms
for mathematical operations and convolutions and a robust mechanism for error control during numerical computations.

Once \mrcpp{} was available, we realized it would be useful to make its functionality easily accessible. 
For this purpose we turned our attention to Python, which has become a \emph{de facto} standard in the scientific community, widely taught to science students and used extensively in research. Its high-level syntax and rich 
ecosystem, including libraries such as \numpy{}\cite{numpy}, \scipy{}\cite{scipy} and \matplotlib{}\cite{matplotlib}, make it a powerful tool for scientific computing. Moreover, Python's integration with Jupyter Notebooks\cite{jupyter} has 
facilitated interactive and reproducible research. This is in contrast to traditional quantum chemistry codes, which are implemented in lower-level languages (Fortran, C, or C++), and can pose significant barriers for non-expert programmers.

That led us to the development of \vampyr{} (Very Accurate Multiwavelets Python Routines) \cite{vampyr_github}, which builds on the \mrcpp{} capabilities, by making them available in the high-level programming framework provided by Python, thereby simplifying the processes of code development, verification, and exploration.  The purpose of \vampyr{} is to allow a broader audience to make use of these tools while maintaining the original power and efficiency of the \mrcpp{} library. This includes inheriting \mrcpp{}'s parallelization which employs the multi-platform 
shared-memory parallel programming model provided by OpenMP.\cite{Van_der_Pas2017-wq, Mattson2019-gl}

\section{Multiwavelets}
\label{sec:mra}

Tha foundations of Multiresolution Analysis (MRA) have been laid in the early '80 thanks to the pioneering work of Meyer, Daubechies, Strömberg and others, \cite{Haar_1910,Stromberg_1981,Battle_1987,Lemarie_1988,Meyer_1985,Meyer_1986,Steffen_1993,Coifman_1994} who defined for the first time the concepts of wavelet functions and nested wavelet spaces. We here provide a short survey by focusing on a particular kind of wavelets called \emph{multiwavelets} \cite{Mallat_1989a,Mallat_1989b,Sweldens_1998,alpert,abgv}.

\subsection{Multiresolution Analysis}

The construction of a Multiresolution Analysis (MRA) starts by considering a basis of generating functions, called \emph{scaling} and \emph{wavelet} functions. We will here limit ourselves to the specific case of \emph{multiwavelets}, referring to the literature for a wider presentation of the subject\cite{Haar_1910,Stromberg_1981,Battle_1987,Lemarie_1988,Meyer_1985,Meyer_1986,Steffen_1993,Coifman_1994}.
The most common choice is the set of $k+1$ polynomials up to order $k$ in the unit interval, such as the Legendre polynomials or interpolating Lagrange polynomials \cite{alpert,abgv}.
The original scaling function can be translated and dilated:
\begin{align}
\label{eq:scaling-basis}
\varphi_{j, l}^n(x) = 2^{n/2} \varphi_i(2^n x - l)
\end{align}

In the equation above $\varphi$ are the generating polynomials, $j$ is the polynomial index ($j=0, 1, \ldots k)$, $n$ is the \emph{scale} index and $l$ is the \emph{translation} index. For a given scale $n$ the functions $\varphi^n$ constitute a \emph{scaling space} $V^n$.
By construction, each interval $V^n_l$ in $V^n$ splits into two sub-intervals in $V^{n+1}$: $V^{n+1}_{2l}$ and $V^{n+1}_{2l+1}$, doubling the basis with every $n$ along the ladder.
This establishes the nested structure \(V^{n} \subset V^{n+1}\) seen in \eqref{eq:mra}. It is also possible to show that this hierarchy is dense in $L^2(\mathbb{R})$.

\begin{align}
\label{eq:mra}
\cdots \subset V^{-1} \subset V^{0} \subset V^{1} \subset \cdots \subset V^n \subset \cdots
\end{align}


The \emph{wavelet space} $W^n$ is defined as the orthogonal complement of $V^n$ with respect to $V^{n+1}$:
\begin{align}
\label{eq:wavelet-space}
W^{n} = V^{n+1} \ominus V^{n}
\end{align}
The wavelet functions are also supported on the same disjoint intervals on the real line as the corresponding scaling functions. Dilation and translation relationships hold for wavelet functions as well: \begin{align}
\label{eq:wavelet-basis}
\psi_{i, l}^n(x) = 2^{n/2} \psi_i(2^n x - l)
\end{align}
By construction, the wavelet functions \(\psi_i\) are orthogonal to the polynomials in the corresponding scaling space. This property, known as ``vanishing moments'', is key in achieving fast algorithms for various operations with multiwavelets. The construction of the multiwavelet functions in not unique (each $W^n_l$ spans a $k+1$ dimensional space and any orthonormal basis in this space is a legitimate choice). We follow the construction presented by Alpert \cite{alpert}, which both maximizes the number of vanishing moments and divides the functions in symmetric and antisymmetric ones.

We finally note that Equation \eqref{eq:wavelet-space} leads to the following telescopic series:
\begin{align}
\label{eq:multires-space}
V^{N} = V^0 \oplus W^0 \oplus W^1 \oplus \cdots \oplus W^{N-1}
\end{align}

\subsection{Multiwavelet transform}
\label{sec:two-scale}

The ladder of spaces structure which is summarized in Eq.~\eqref{eq:wavelet-space} and Eq.~\eqref{eq:multires-space} implies that it is possible to span $V^{n+1}$ either through the scaling functions $\scaling^{n+1}$ or through the scaling functions $\scaling^n$ and the wavelet functions $\wavelet^n$. The basis set transformation between these two bases is known as the \emph{multiwavelet transform} or \emph{two-scale filter relationship}. The unitary transformation matrix is assembled by making use of the four wavelet \emph{filters} $G^{(0)}$, $G^{(1)}$, $H^{(0)}$ and $H^{(1)}$

\begin{equation}
\label{eq:twoscalerelations}
\begin{pmatrix}
\varphi_{l}^{n}\\
\psi_{l}^{n}
\end{pmatrix}=
\begin{pmatrix}
H^{(0)}&H^{(1)} \\
G^{(0)}&G^{(1)}
\end{pmatrix}
\begin{pmatrix}
\varphi_{2l}^{n+1} \\
\varphi_{2l+1}^{n+1}
\end{pmatrix}
\end{equation}

The matrices $H^{(0)}, H^{(1)}, G^{(0)}$, and $G^{(1)}$ are each of dimension $(k+1) \times (k+1)$. See Ref.~\citenum{abgv} for a comprehensive derivation of these matrices.
The operation illustrated in Eq.~\eqref{eq:twoscalerelations} is called \emph{forward wavelet transform}, also known as \emph{wavelet compression}. The inverse of this operation is termed \emph{backward wavelet transform} or \emph{wavelet reconstruction}. The operation is strictly local: the scaling function $\phi_l^n$ and wavelet function $\psi_l^n$ are only connected to $\phi^{n+1}_{2l}$ and $\phi^{n+1}_{2l+1}$.
This structure simplifies numerical algorithms, enabling fast implementations.

\subsection{Function projection onto the multiwavelet space}
\label{subsec:pjf}

The most basic operation in \ac{MRA} is the projection of a function. It generates the representation of an arbitrary function in a MW basis.
Let us consider the scaling space \( V^n \), and the associated projector \( P^n \). The MW projection of a given function $f(x)$ can be obtained as:
\begin{align}
\label{eq:scaling-proj}
f^n(x) = P^n f (x) =  \sum_{l=0}^{2^n - 1} \sum_{j=0}^k s^{n,f}_{j, l} \varphi_{j,l}^n(x),
\end{align}
Here, \( s^{n,f}_{j, l} \) are the \emph{scaling coefficients}, given by:
\begin{align}
\label{eq:scaling-coef}
s^{n, f}_{j, l} = \int f(x) \varphi_{j, l}^n(x) dx.
\end{align}
Similarly, a wavelet projector \( Q^n \) is associated with the wavelet space \( W^n \):
\begin{align}
\label{eq:wavelet-proj}
df^n(x) =  Q^n f (x) = \sum_{l=0}^{2^n - 1} \sum_{j=0}^k w^{n,f}_{j, l} \psi_{j,l}^n(x),
\end{align}
The wavelet coefficients \( w^{n,f}_{j,l} \) can also be obtained as:
\begin{align}
\label{eq:wavelet-coef}
w^{n, f}_{j, l} = \int f(x) \psi_{j, l}^n(x) dx.
\end{align}

\subsection{Adaptive Projection}
\label{subsec:adaptive-proj}

Keeping in mind the projections defined in the previous section, the complete representation of a function in a given \ac{MRA} can be written as follows:
\begin{align}
    P^N f = f^N = f^0 + \sum_{n=0}^{N-1} df^n = (P^0 + \sum_{n=0}^{N-1} Q^n) f
\end{align}
This equation gains special significance when considering that the wavelet coefficients, \( w^n_l \), 
approach zero for smooth functions. The precision of the representation can thus be assessed by inspecting 
the wavelet coefficients at a specific scale \( n \) and translation \( l \):
\begin{align}
    |w^n_l| < \epsilon.
\end{align}
This consideration provides the foundation for an adaptive projection strategy: rather than fixing the projection to a predefined scale \( N \), one can incrementally increase the scale from  $0$ only in those intervals where the wavelet norm is larger than the requested precision. Utilizing the two-scale filter relations, Eq.~\eqref{eq:twoscalerelations}, we can compute the wavelet coefficients and assess their magnitude. If these coefficients meet predefined precision criteria at a specific translation \( l \), that branch of the function representation can be truncated. This truncation effectively focuses computational and data resources only where high precision is necessary. Should the desired precision level not be reached, the refinement process continues in a recursive manner. This recursion can in principle be carried out indefinitely until precision requirements are met. In practice, a maximum allowed scale is set.


\subsection{Operator Application in Multiwavelet Framework -- Non-Standard Form}
\label{subsec:non-standard-form}

The most convenient strategy to apply an operator using \acp{MW} is to construct its \ac{NS} form. Similarly to functions, one can construct the operator projection $T^N = P^N T P^N$. By recalling that for every scale $n$, $P^{n+1} = P^n + Q^n$ one obtains:
\begin{align}
    T^N &= P^N T P^N = (P^{N-1} + Q^{N-1}) T (P^{N-1} + Q^{N-1}) \\ 
        &= Q^{N-1} T Q^{N-1} + Q^{N-1} T P^{N-1} + P^{N-1} T Q^{N-1} + P^{N-1} T P^{N-1} \\
        & = A^{N-1} + B^{N-1} + C^{N-1} + T^{N-1}
\end{align}
where for each $n$ the following definitions are used:
\begin{align}
    T^n = P^n T P^n, &&
    A^n = Q^n T Q^n, \\
    B^n = Q^n T P^n, &&
    C^n = P^n T Q^n.
\end{align}

This structure is the operator analogue of the two-scale relationship for functions and it can be extended telescopically to obtain:
\begin{align}
    T^N = T^0 + \sum_{n=0}^{N} \left( A^n + B^n + C^n \right).
    \label{eq:non-standard-form}
\end{align}
\( A^n \), \( B^n \), and \( C^n \) components exhibit sparsity properties, similar to what is observed for the wavelet coefficients of function representations, and are therefore leading to fast, and often linearly scaling algorithms, for any arbitrary predefined precision. The purely scaling part $T^0$ of the operator is only required at the coarsest scale, where only a handful of grid points are present. Another important feature of the \ac{NS} form is the absence of coupling between different scales, which allows to preserve the adaptive precision of the representation on the one hand, and independent or asynchronous operator application across different scales, which boosts computational efficiency, on the other hand. For additional details about the application of operators the non-standard form \cite{BEYLKIN2008354}, the reader is referred to the literature on the subject~\cite{Beylkin1991-dd,abgv}.

\section{\vampyr{} }

\begin{figure}[htb]
    \centering
    \includegraphics[width=0.5\textwidth]{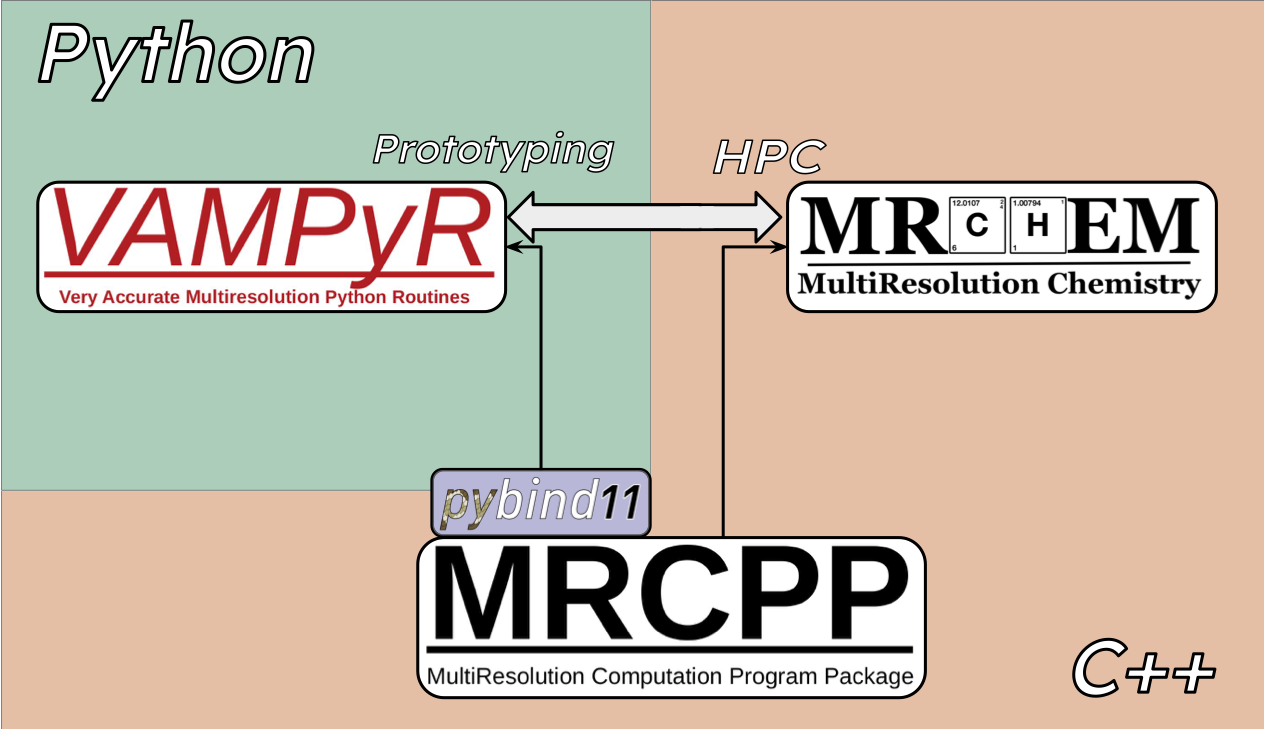}
    \caption{Connections between \mrchem{}, \mrcpp{} and \vampyr{}. \mrcpp{} implements a high-performance \ac{MRA} framework with \acp{MW}. It implements the representation of functions and operators as well as fast algorithms for operator application. \mrchem{} implements electronic structure methods and uses \mrcpp{} for the underlying mathematical operations. \vampyr{} imports features from \mrcpp{} using \pybind{} and brings intuitive design and easy prototyping through Python. It is used as a Python library can therefore be interfaced other quantum chemistry software. \vampyr{} doesn't include functionality from \mrchem{}, at the current stage.
    All software packages are available on GitHub, under the \mrchemsoft{} organization\cite{mrchemsoft_github}.
    }
    \label{fig:mrcpp-pybind11-vampyr}
\end{figure}
In the architecture of \vampyr{}, \mrcpp{} serves as the foundational layer, integrated into Python through \pybind{}, as illustrated in Figure \ref{fig:mrcpp-pybind11-vampyr}. This setup enables seamless interoperability between Python and C++, allowing \pybind{} to automatically convert many standard C++ types to their Python equivalents and vice versa. This leads to a more Pythonic and natural interface to the \mrcpp{} codebase.

For the development of \vampyr{}, we opted for \pybind{} as our binding framework, chiefly for its robustness, maintainability, and alignment with our development objectives. \pybind{} is a lightweight, header-only library that utilizes modern C++ standards to infer type information, thus streamlining the binding process and enhancing overall code quality.

To facilitate deployment, multiple installation options are available for \vampyr{} and \mrcpp{}. The source code for both is openly accessible and distributed under the LGPLv3 license on GitHub, specifically within the \mrchemsoft{} organization\cite{mrchemsoft_github}. For users who prefer a simplified installation procedure, binary packages are also provided through the Conda package manager, which is compatible with various Linux and macOS architectures\cite{mrcpp_conda-forge, vampyr_conda-forge}. Each new release triggers an automated build process that uploads the binary packages to the Conda Forge channel on \url{anaconda.org}, thereby easing the installation process and incorporating all requisite dependencies. The packages are designed to be compatible with current Python versions, ranging from 3.8 to 3.11. The installation process using Conda is straightforward:
\begin{minted}{bash}
    conda install vampyr -c conda-forge
\end{minted}

\subsection{Implementation}



In \mrcpp{}, C++ template classes and functions are utilized to provide abstraction over the dimensionality of the simulation box. These templates enable the generic implementation of data structures and algorithms for problems in 
$D$ dimensions.\footnote{It should be noted that some performance-oriented functionalities, as well as specific operators like Helmholtz and Poisson, are currently exclusive to 3-dimensional problems. The extensions to lower or higher dimensions are possible, but are currently not implemented.} The specialization for 1-, 2-, and 3-dimensional problems is performed at compile-time, thereby eliminating any impact on runtime performance.\cite{Vandevoorde2018-ky}

In Python, native template constructs are absent, presenting a challenge for dimension-specific specialization. In \vampyr{}, we emulate \mrcpp{}'s generic approach by implementing dimension-dependent bindings using \pybind{}. Specifically, our binding code incorporates template classes and functions similar to those in \mrcpp{}. \mrcpp{}'s 1-, 2-, and 3-dimensional template classes and functions are directly bound to their corresponding dimensions in \vampyr{}, as demonstrated in Listing \ref{lst:import}. Through this approach, \vampyr{} keeps the flexible design of \mrcpp{}, despite Python's limitations in handling generic datatypes.

\begin{listing}[!ht]
\caption{Importing dimensional-dependent bindings from \vampyr{}.}
\begin{minted}{python}
from vampyr import vampyr1d
from vampyr import vampyr2d
from vampyr import vampyr3d
\end{minted}
\label{lst:import}
\end{listing}

While the \texttt{MultiResolutionAnalysis} (see section \ref{sec:mra-vampyr}) and \texttt{FunctionTree} (see section \ref{subsec:pjf-vampyr}) classes in \vampyr{} are inherited from 
\mrcpp{}, there are noteworthy distinctions between the vanilla C++ and the corresponding Python classes. These 
differences are introduced to \emph{enhance} the Python version of the \mrcpp{} classes with so call \emph{syntactic sugar}, which improves the user experience and leads to faster prototyping.
Among these enhancements are the overload of various \emph{dunder} (double underscore, also known as \emph{magic}) 
methods that have been added to the classes to extend their functionalities and make them more Pythonic.

These \emph{dunder} (or magic) methods enable the use of Python's built-in operators on \texttt{FunctionTree} objects. For instance, the
addition (\verb|__add__|) and multiplication (\verb|__mul__|) operators allow for the direct use of \texttt{+} and \texttt{*} 
operators with \texttt{FunctionTree} objects. This allows 
to write more intuitive code, which is easier to both read and write. 

As an example, consider the addition of two \texttt{FunctionTree} objects, which be expressed in natural Python syntax (Listing \ref{lst:simple_syntax}), thanks to the binding code in Listing \ref{lst:binding_aritmetics}.

\begin{listing}[!ht]
\caption{Syntax sugar for \texttt{FunctionTree} addition in action.}
\label{lst:simple_syntax}
\begin{minted}{python}
f_tree1 = vp.FunctionTree(MRA)
f_tree2 = vp.FunctionTree(MRA)
f_tree_sum = f_tree1 + f_tree2
\end{minted}
\end{listing}

\begin{listing}[!ht]
\caption{C++ binding for operator \texttt{+}.}
\label{lst:binding_aritmetics}
\begin{minted}{c++}
// allocate new FunctionTree for the output
auto out = std::make_unique<FunctionTree<D>>(inp_a->getMRA());
// arrange input operands in vector of summands
FunctionTreeVector<D> vec;
vec.push_back({1.0, inp_a});
vec.push_back({1.0, inp_b});
// build grid for output value based on summands
build_grid(*out, vec);
// perform summation with no additional refinements
add(-1.0, *out, vec);
// return sum of FunctionTree objects
return out;
\end{minted}
\end{listing}

The Python code in Listing \ref{lst:simple_syntax} can be equivalently written in the de-sugared form in Listing \ref{lst:desugared}. Although the Listing \ref{lst:desugared} is more involved than simply applying an arithmetic operator, it does provide more flexibility and control which might be necessary in some specific applications.

\begin{listing}[!ht]
\caption{De-sugared syntax for \texttt{FunctionTree} addition using the \texttt{advanced} bindings submodule.}
\label{lst:desugared}
\begin{minted}{python}
from vampyr import vampyr3d as vp

f_tree1 = vp.FunctionTree(MRA)
f_tree2 = vp.FunctionTree(MRA)

f_tree_sum = vp.FunctionTree(mra)

vp.advanced.add(-1.0, f_tree_sum, 1.0, f_tree1, 1.0, f_tree2)
\end{minted}
\end{listing}

\section{Mathematics with \vampyr{}}
\label{sec:math-vampyr}

In this section we display some of the theoretical concepts discussed in Section \ref{sec:mra} with
practical examples using \vampyr{}.


\subsection{The \texttt{MultiResolutionAnalysis} Object in \vampyr{}}
\label{sec:mra-vampyr}
The foundation of \vampyr{}'s internal operations is built upon the \texttt{MultiResolutionAnalysis} object. This object encapsulates essential 
information about the physical space being modeled, along with configurations for the scaling and wavelet basis functions as defined in the theory 
of \ac{MRA} (see Section~\ref{sec:mra}).

\paragraph{Standard Usage}
For users seeking a straightforward implementation, \vampyr{} provides a ``Standard Usage'' option. In this approach, users are only required to specify the box size and polynomial order. The remaining parameters, such as the choice of interpolating polynomials for the scaling basis, are automatically configured by the library. Listing~\ref{lst:standard_usage} illustrate how to create an \ac{MRA} object either specifying the start and end points, or the length: the former has length $2L$ whereas the latter starts in $0$ and ends in $L$.
\begin{listing}[!ht]
\caption{Standard usage of the \ac{MRA} object.}\label{lst:standard_usage}
\begin{minted}{python}
# For a unit cell centered at the origin
MRA = vp.MultiResolutionAnalysis(box=[-L,L], order=k)
# For a unit cell with left corner at the origin
MRA = vp.MultiResolutionAnalysis(box=[L], order=k)
\end{minted}
\end{listing}
\paragraph{Advanced Usage}
For users requiring more control over the computational setup, \vampyr{} offers an ``Advanced Usage'' approach. This method allows for the customization of various parameters, including the world box size and the specific type of scaling basis, such as Legendre polynomials. This level of customization grants the user full control over the basis configuration, as demonstrated in Listing~\ref{lst:advanced_usage}.
\begin{listing}[!ht]
\caption{Advanced usage of the MRA object.}\label{lst:advanced_usage}
\begin{minted}{python}
# Construct scaling basis
scaling_basis = InterpolatingBasis(order) # or LegendreBasis(order)
# Define world box
world_box = BoundingBox(scale=0, corner=[x0, y0, z0],
            nboxes=[Nx, Ny, Nz], scaling=[a, b, c]) 
            # in 3D, for 1D and 2D replace list of size 3 with list of size 1 or 2
# Construct MRA
MRA = vp.MultiResolutionAnalysis(box=world_box, basis=scaling_basis)
\end{minted}
\end{listing}
This flexibility allows users to choose the level of interaction with the \ac{MRA} object based on their specific needs and expertise.

\subsection{Function Projectors}
\label{subsec:pjf-vampyr}

In \vampyr{}, functions are represented using a tree-based data structure: the \texttt{FunctionTree} object. This structure naturally arises from the \ac{MRA} framework outlined in Section~\ref{subsec:pjf}.

\paragraph{The Tree}
The \texttt{FunctionTree} object consists of interconnected nodes organized hierarchically. 
The root node\footnote{In general, one root node is sufficient, but it is possible to specify a domain containing more than one root node. This option can be useful in the multidimensional case, if one wants a domain which had different sizes in the different directions.} represents the entire physical domain, and each non-terminal node or \emph{branch} begets \(2^d\) child nodes while connecting to a single parent node. 
Terminal nodes, or \emph{leaves}, possess a parent but have no children. 
Here, \(d\) represents the dimensionality of the system. \vampyr{} currently supports $d=1, 2, 3$.

\paragraph{The Node}
Each node corresponds to a \(d\)-dimensional box within the physical domain and encapsulates information about a specific set of scaling and wavelet functions defined therein. 
Specifically, each node indexed by scale $n$ and translation vector \(\mathbf{l} = (l_1, l_2, \ldots, l_d)\) stores \((k+1)^d\) scaling coefficients and \((2^d-1)(k+1)^d\) wavelet coefficients.
Here the indices $n$ and $\mathbf{l}$ dictate the node's spatial size and position, respectively. 

\subsection{Scaling and Wavelet projectors in \vampyr{}}

To facilitate the implementation of function projections into scaling and wavelet spaces, \vampyr{} provides the \texttt{vp.ScalingProjector} 
and \texttt{vp.WaveletProjector} classes. In Listing~\ref{lst:1d_proj}, we define a lambda function  \texttt{f}, whose body can be any valid Python expression.\footnote{We note that the projection in the \ac{MRA} of arbitrary Python functions is executed \emph{serially} in \vampyr{}, since it is not in general threadsafe to release Python's so-called global interpreter lock (GIL) to exploit \mrcpp{} OpenMP parallelization.} This illustrates how to project a function using the \texttt{vp.ScalingProjector} 
and \texttt{vp.WaveletProjector} classes. 
Instances \texttt{P\_n} and \texttt{Q\_n} are created to perform the projection into the scaling and wavelet spaces. 
The resulting representations, denoted \texttt{f\_n} and \texttt{df\_n}, are fully described by their mathematical definitions in Equations~\eqref{eq:scaling-proj} and~\eqref{eq:wavelet-proj}, respectively.

\begin{listing}[!ht]
\caption{Function Projection in \vampyr{} by Scaling and Wavelet projectors.}\label{lst:1d_proj}
\begin{minted}{python}
f = lambda x: <analytic function>
P_n = vp.ScalingProjector(MRA, scale=n)
Q_n = vp.WaveletProjector(MRA, scale=n)
f_n = P_n(f)
df_n = Q_n(f)
\end{minted}
\end{listing}

\begin{figure}[hb]
  \subfigure[\label{fig:scaling_proj0}]{%
    \includegraphics[width=0.45\textwidth]{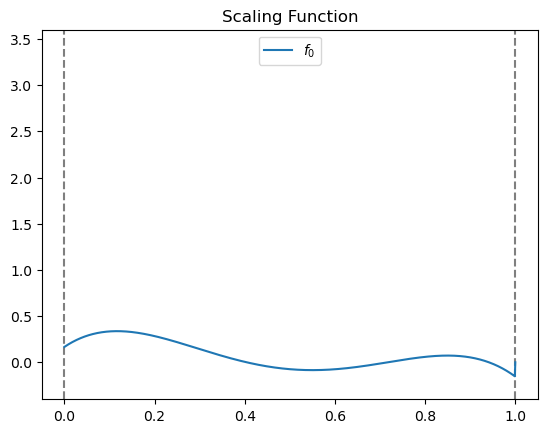}
  }
  \hfill
  \subfigure[\label{fig:wavelet_proj0}]{ 
    \includegraphics[width=0.45\textwidth]{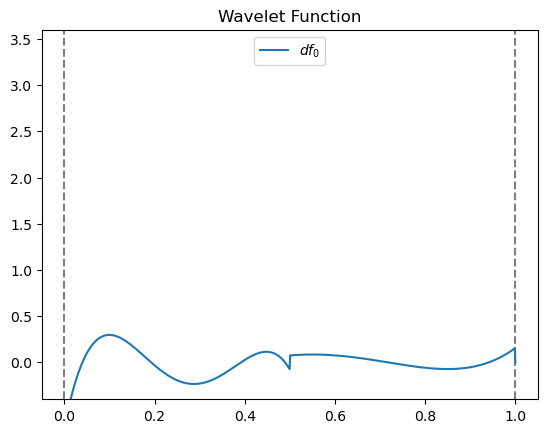}
  }
  \caption{Projection of a Slater function at root scale. Left: Projection onto scaling function space $V^0_4$ with vertical lines marking the physical space spanned by $k+1$ scaling functions. Right: Projection onto wavelet function space $W^0_4$, where vertical lines indicate the physical space spanned by $k+1$ wavelet functions.}
  \label{fig:proj0}
\end{figure}

\begin{figure}[!hb]
  \subfigure[\label{fig:scaling_proj4}]{%
    \includegraphics[width=0.45\textwidth]{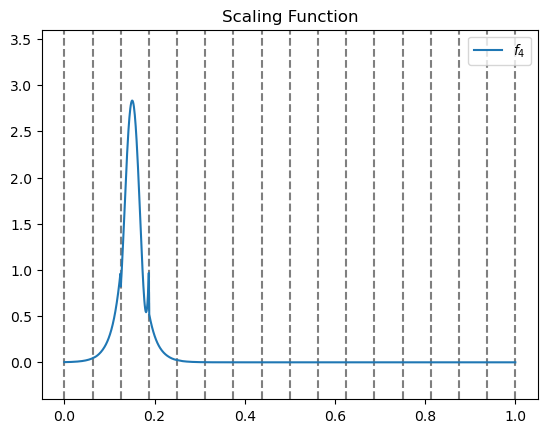}
  }
  \hfill
  \subfigure[\label{fig:wavelet_proj4}]{ 
    \includegraphics[width=0.45\textwidth]{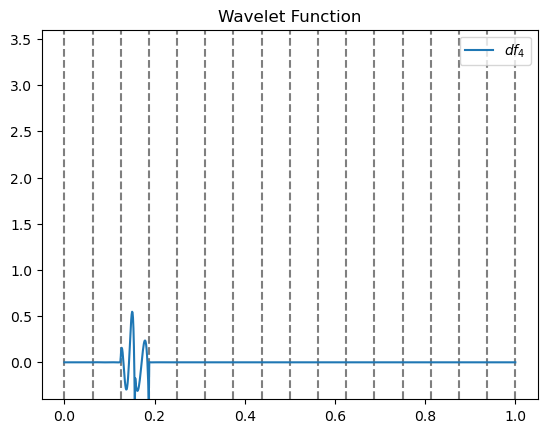}
  }
  \caption{Projection of a Slater function at the 4th scale. Left: Projection onto scaling function space $V^4$, with vertical lines delineating the physical space spanned by $k+1$ scaling functions. Right: Projection onto wavelet function space $W^4$, where vertical lines mark the physical space spanned by $k+1$ wavelet functions.}
  \label{fig:proj4}
\end{figure}

Figures~\ref{fig:proj0} and~\ref{fig:proj4} provide visual insights into the projection of an exponential (Slater) function, $e^{- a |r|}$, onto different scales. Each figure consists of two subfigures: the left-hand side displays the scaling function space (Figures~\ref{fig:scaling_proj0} and~\ref{fig:scaling_proj4}), while the right-hand side illustrates the wavelet function space (Figures~\ref{fig:wavelet_proj0} and~\ref{fig:wavelet_proj4}).

In Figure~\ref{fig:proj0}, the projection onto the root scale is far from the target, and the large wavelet part indicates that a representation at a finer scale is mandatory. The vertical lines in both subfigures delineate the physical space spanned by \(k+1\) scaling and wavelet functions. In Figure~\ref{fig:proj4}, 
the exponential (Slater) function is projected onto the 4th scale. The representation is now close to its target function: it is however not as sharp as the original
exponential (Slater) function, and the wavelet projection indicates that in the cusp region, a finer representation would be required to attain high precision. 


\subsection{Adaptive Projection in \vampyr{}}

\begin{figure}[!htb]
    \centering
    \includegraphics[width=.7\textwidth]{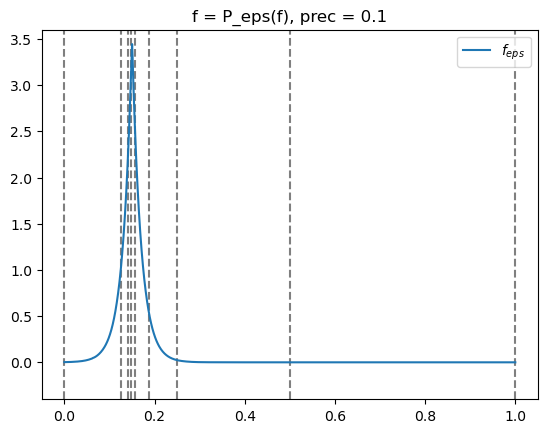}
    \caption{Adaptive projection of the Slater function with \texttt{prec} = \(1.0 \times 10^{-1}\). The non-uniform vertical lines indicate the physical space spanned by the adaptive basis functions, underscoring the method's efficiency and precision.}
    \label{fig:proj_prec1}
\end{figure}

Building on the mathematical framework presented in Section~\ref{subsec:adaptive-proj}, we cover here the practical aspects of adaptive projection within \vampyr{}. Unlike the fixed-scale projection illustrated in Listing~\ref{lst:1d_proj}, adaptive projection makes use of the \texttt{prec} keyword to refine the function's representation, as shown in Listing~\ref{lst:adaptive_proj}.
\begin{listing}
\caption{Adaptive Function Projection in \vampyr{}.}
\label{lst:adaptive_proj}
\begin{minted}{python}
P_eps = vp.ScalingProjector(MRA, prec=epsilon)
f_eps = P_eps(f)
\end{minted}
\end{listing}
The visual comparison in Figure~\ref{fig:proj_prec1} between the adaptive and fixed-scale projections (previously shown in Figures~\ref{fig:proj0} and~\ref{fig:proj4}) reveals the power of adaptivity: a representation which is at the same time more compact as well as more precise is achieved. As Figure~\ref{fig:proj_prec1} shows, nodes at finer and finer resolution are created only where it is necessary to represent the sharp features of the function. This strategy is also computationally more efficient as it guarantees that resources are used where and when needed, based on precision requirements.

\subsection{Arithmetic Operations in FunctionTrees}

In \vampyr{}, arithmetic operations on \texttt{FunctionTree} objects are intuitive and flexible. Standard Python operators like \(+\), \(-\), \(*\), \(/\), and \(**\) are employed for these elementary operations.
For an example see Listing~\ref{lst:arith_ops}. where we have 
examples of usage for both standard arithmetic operations between \texttt{FunctionTree}s and between \texttt{FunctionTree}s and scalars. In these examples, \( f \) and \( g \) are instances of \texttt{FunctionTree}, and \( a \) is a scalar. These operations provide an efficient and user-friendly way to manipulate \texttt{FunctionTree} objects in \vampyr.

\begin{listing}[!ht]
\caption{Basic Arithmetic Operations in \vampyr.}\label{lst:arith_ops}
\begin{minted}{python}
# Arithmetics with scalars
a_mult_f = a * f  # Scalar on the left
f_mult_a = f * a  # Scalar on the right
f_div_a = f / a   # Division by scalar
f_pow_a = f ** a  # Power operator

# Arithmetics between FunctionTrees
f_mult_g = f * g  # Multiplication operator
f_add_g = f + g   # Addition operator
f_sub_g = f - g   # Subtraction operator

# In-place arithmetics
f *= a  # Multiplication with scalar
f *= g  # Multiplication with function
f += g  # Addition with function
f -= g  # Subtraction with function
\end{minted}
\end{listing}


\subsection{Vectorizing \texttt{FunctionTree}s with \numpy}
\label{sec:vectorizing-numpy}

The integration between the overloaded arithmetic methods in \vampyr{} and \numpy's flexible data structures provides a range of computational advantages. Among these advantages, the support for \textbf{multi-dimensional arrays}, \textbf{broadcasting}, and \textbf{linear 
algebra operations} stand out. Although \numpy{} is primarily optimized for integers and floating-point numbers, its data structures can be 
extended to accommodate custom objects such as \functiontree s. This capability enables the incorporation of \functiontree{}s within \numpy{} 
arrays, thus allowing for the full suite of \numpy's array-oriented computing features.
For illustration, consider the example in Listing \ref{lst:ft_numpy_example}, where we construct a matrix of \functiontree{}s within a \numpy{} array and execute various matrix operations.

\begin{listing}[!ht]
\caption{Performing matrix operations using function trees in a \numpy{} array.}\label{lst:ft_numpy_example}
\begin{minted}{python}
# Creation of a matrix containing function trees
matrix_a = np.array([[a11_tree, a12_tree], [a21_tree, a22_tree]])
matrix_b = np.array([[b11_tree, b12_tree], [b21_tree, b22_tree]])

# Performing matrix multiplication
matrix_mul = matrix_a @ matrix_b

# Performing addition and subtraction
matrix_sum = matrix_a + matrix_b
matrix_neg = matrix_a - matrix_b

# Multiplication by a scalar
matrix_mul_scalar = 2 * matrix_a * matrix_b

# Demonstrating broadcasting by adding a single function tree to all elements
matrix_a_bcast_add = matrix_a + single_tree
\end{minted}
\end{listing}

As demonstrated in Listing \ref{lst:ft_numpy_example}, the integration between \vampyr's \functiontree{} objects and \numpy{} arrays simplifies the implementation of complex mathematical operations. Through \numpy's multi-dimensional arrays, it is possible to organize and manipulate arrays of function trees efficiently. Broadcasting allows for the addition of a single \functiontree{} (or scalar) to an entire array of \functiontree s. Furthermore, \numpy's built-in linear algebra functions, such as the matrix multiplication operator \texttt{@}, can be used for matrix operations. This approach improves the readability of the code, as will be seen in Section \ref{sec:vampyrSCF}.

\subsection{Derivative Operators in \vampyr{}}
\label{sec:deriv-operators}
Handling derivatives with \acp{MW} is not straightforward: traditional derivative operators are not applicable. However, \vampyr{} offers two specialized operators to address this issue, each tailored for specific requirements concerning the continuity of the 
function and the desired order of the derivative. These methods are based on the works by \citeauthor{abgv}, giving the \texttt{ABGVOperator},\cite{abgv} and \citeauthor{anderson2019derivatives}, leading to the \texttt{BSOperator}.\cite{anderson2019derivatives}

Both the \verb|ABGVOperator| and the \verb|BSOperator| apply the operators following the \ac{NS} form as described in Subsection \ref{subsec:non-standard-form}, allowing for efficient and accurate calculations:
\begin{description}
    \item[\texttt{ABGVOperator}] Designed for piece-wise continuous functions, this operator provides a weak formulation for first-order derivatives. It is primarily suitable for handling first-order derivatives and offers a balance between accuracy and computational efficiency. Usage is demonstrated in Listing \ref{lst:abgv_example}.
    \item[\texttt{BSOperator}] This operator is more versatile but it works best for smooth continuous functions. It accommodates higher-order derivatives by transforming the function onto a B-spline basis before differentiation. See Listing \ref{lst:bs_example} for a practical example.
\end{description}

\begin{listing}[!ht]
\caption{First-order derivatives using \texttt{ABGVOperator}.}\label{lst:abgv_example}
\begin{minted}{python}
D = vp.ABGVOperator(MRA)

# x, y, z derivatives using ABGVOperator
f_x = D(f, 0)  
f_y = D(f, 1)  
f_z = D(f, 2)  
\end{minted}
\end{listing}

\begin{listing}[!ht]
\caption{First- and second-order derivatives using \texttt{BSOperator}.}\label{lst:bs_example}
\begin{minted}{python}
D1 = vp.BSOperator(MRA, 1) # First-order derivative operator
D2 = vp.BSOperator(MRA, 2) # Second-order derivative operator

# First-order derivatives
f_x = D1(f, 0)
f_y = D1(f, 1)
f_z = D1(f, 2)

# Second-order derivatives
f_xx = D2(f, 0)
f_yy = D2(f, 1)
f_zz = D2(f, 2)
\end{minted}
\end{listing}

\subsection{Convolution Operators in \vampyr{}}
\label{sec:conv-operators}

Convolution operators are essential tools for solving integral forms of key equations in Quantum Chemistry, such as the integral formulation of the \ac{HF} and \ac{KS} equations or the Poisson equation used in \ac{SCRF} models. Both equations can be recast from the more familiar differential form, to an equivalent integral form by making use of the \ac{BSH} kernel (see Appendix \ref{sec:HFappendix}) and the Poisson kernel. The key step in both problems is the convolution of the corresponding Green's function kernel with a test function:
\begin{equation}\label{eq:convolution}
g(\mathbf{x}) = \int K(\mathbf{x}, \mathbf{y}) f(\mathbf{y}) \text{d}\mathbf{y}.
\end{equation}
The two kernels are numerically implemented in the \mrcpp{} library and available in \vampyr{}. They are constructed as a sum of Gaussians which represents the numerical quadrature of the integral of a superexponentially decaying function\cite{frediani2013fully} which depends on the distance $d =  | \mathbf{x} - \mathbf{y} |$. 
\begin{align}
K(\mathbf{x}, \mathbf{y}) = \frac{e^{- \mu | \mathbf{x} - \mathbf{y} |}}{| \mathbf{x} - \mathbf{y} |} \approx \sum_{i=1}^M \alpha_i \exp(-\beta_i | \mathbf{x} - \mathbf{y}|^2),
\end{align}
In the expression above, \(\mu = 0\) yields the Poisson kernel, and \(\mu > 0\) corresponds to the bound-state Helmholtz kernel.


\begin{listing}[!ht]
\caption{Creating and applying Poisson and Helmholtz operators in \vampyr{}.}\label{lst:poisson_helmholtz_operator}
\begin{minted}{python}
P_oper = vp.PoissonOperator(mra, prec)
H_oper = vp.HelmholtzOperator(mra, mu, prec)

g_tree = P_oper(f_tree)
g_tree = H_oper(f_tree)
\end{minted}
\end{listing}

\begin{figure}
  \subfigure[Original image projected onto the Multi-Resolution Analysis (MRA).\label{fig:img-mra}]{%
    \includegraphics[width=0.4\textwidth]{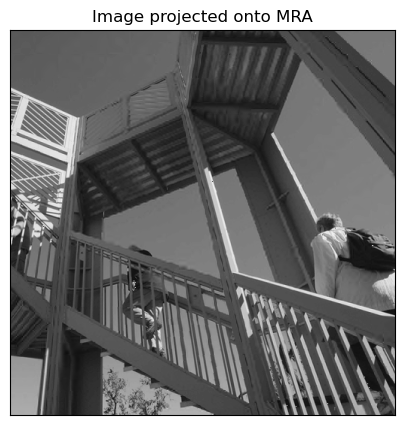}
  }
  \hfill
  \subfigure[Resulting image after performing convolution with a Gaussian Kernel, leading to smoothing or blurring of the original image.\label{fig:img-conv}]{%
    \includegraphics[width=0.4\textwidth]{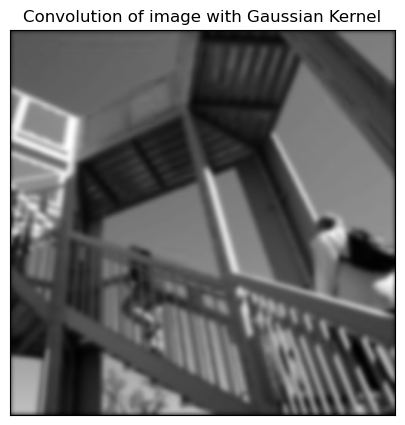}
  }
\caption{Illustration of convolution process in \vampyr{}.}
\label{fig:image}
\end{figure}

A more general method for constructing convolution operators based on Gaussian expansions is also available in \vampyr{}. This approach is documented with an example in Listing \ref{lst:convolution-operator} and an unusual application is illustrated in Figure \ref{fig:image} by blurring an image with a Gaussian convolution kernel, although \vampyr{} is not generally optimized for such tasks.

\begin{listing}[!ht]
\caption{Construction of a Custom Convolution Operator in \vampyr{}.}\label{lst:convolution-operator}
\begin{minted}{python}
# Create a 1D GaussExp object, serving as a container for Gaussians
kernel = vp1.GaussExp()
# Append a Gaussian function to the kernel
kernel.append(vp1.GaussFunc(beta, alpha))
# Construct the convolution operator
T = vp.ConvolutionOperator(mra, kernel, prec)
# Apply the operator to a function tree
g_tree = T(f_tree)
\end{minted}
\end{listing}

\section{Four little pieces of quantum chemistry}
\label{sec:vampyrSCF}

One of the main technical challenges of quantum chemistry is the large gap between the equations describing the physical nature of the system and their practical realization in a working code.
This gap arises because the equations in their original form are in general too complicated to be solved.
To achieve equations which have a manageable computational cost, it has so far been necessary to make use of representations which on the one hand lead to practical numerical implementation, but on the other hand obfuscate the original physical equations.

We will here present \emph{four little pieces} of quantum chemistry where we show how the \ac{MW} representation of functions and operators in \mrcpp{} and the simple Python interface provided by \vampyr{} can close this semantic gap.

For each of the four examples presented, we will highlight how \vampyr{} has been employed to obtain working implementations that closely resembles the theoretical framework. The underlying equations will here be presented briefly to highlight their correspondence with the code. We refer to separate appendices, or previously published work, for a more in-depth exposition of the theory.

All examples are available as Jupyter notebooks in a separate GitHub repository\cite{vampyr_coven}, openly accessible under the CC-BY-4.0 license.

\subsection{The Hartree--Fock Equations}
The Hartree-Fock equations\cite{jensen2017introduction} can be concisely written as:
\begin{equation}
    \hat{F} \orbital_i = \sum_j F_{ij} \orbital_j
\end{equation}
where $\hat{F}$ is the Fock operator, $\orbital_i$ is an occupied orbital and $F_{ij}$ are the Fock matrix elements in the basis of the occupied orbitals. 

Traditionally, a basis-set representation is introduced and most of the successive discussion regards methods to achieve a good performance and a sufficient precision. A \ac{MW} representation avoids this complication and solves the equations \emph{directly}. As shown in Appendix \ref{sec:HFappendix}, this is achieved by recasting the problem as an integral equation:
\begin{equation}
\label{eq:scf}
\tilde{\orbital}_i^{n+1} = -2 \hat{G}^{\mu^n_i}\left[\left(V_N + 2J^n -K^n\right) \orbital_i^n - \sum_{j \neq i}F^n_{ij}\orbital_j^n \right]
\end{equation}
In the above equation, $\hat{G}^{\mu_i^n}$ is the bound-state Helmholtz operator, $V_N$, $J$ and $K$ are the nuclear, Coulomb and exchange operator, respectively and we have restricted ourselves to closed-shell systems. $i$ is the orbital index and $n$ represents the iteration index and is displayed for all quantities that are iteration-dependent. The resulting orbitals $\tilde{\orbital}_i$ are not normalized (here indicated with a $\tilde{}$ superscript) and an orthonormalization step must then be performed. The complete algorithm can be found in the accompanying Python notebook \cite{vampyr_coven} and 
includes a L\"owdin orthogonalization of the orbitals, ensuring the desired orthogonality condition.
We base our implementation on these equations. The \ac{SCF} equations \eqref{eq:scf} are implemented in 
Python, as demonstrated in Listing \ref{lst:scf_python}.
Thanks to the \vampyr{} interface the code is now directly corresponding to the mathematical formalism without any intermediate representation.

\begin{listing}[!ht]
\caption{Python implementation of the SCF equation.}
\label{lst:scf_python}
\begin{minted}{python}
VPhi = V_nuc(Phi_n) + 2*J_n(Phi_n) - K_n(Phi_n)
Phi_np1 = -2*G(VPhi + (Lambda_n - F_n) @ Phi_n)
\end{minted}
\end{listing}

To achieve an even more compact structure, the set of orbitals can be collected in a \numpy{} vector. This requires that the implementation of operators in the code accepts a vector of orbitals as input.
The operators \verb|V_nuc|, \verb|J_n|, and \verb|K_n| (defined in equations  \eqref{eq:one-electron-op}, \eqref{eq:couloumb_oper}, and \eqref{eq:exchange_oper}, respectively) are implemented in a vectorized manner using \numpy{}. For example, the \texttt{CouloumbOperator} class is presented in Listing \ref{lst:coulomb_operator}.
\begin{listing}[!ht]
\caption{Python implementation of the Couloumb operator.}\label{lst:coulomb_operator}
\begin{minted}{python}
class CouloumbOperator():
    def __init__(self, mra, Psi, prec):
        self.mra = mra
        self.Psi = Psi
        self.prec = prec
        self.poisson = vp.PoissonOperator(mra=mra, prec=self.prec)
        self.potential = None
        self.setup()
    def setup(self):
        rho = self.Psi[0]**2
        for i in range(1, len(self.Psi)):
            rho += self.Psi[i]**2
        rho.crop(self.prec)
        self.potential = (4.0*np.pi)*self.poisson(rho).crop(self.prec)
    def __call__(self, Phi):
        return np.array([(self.potential*phi).crop(self.prec) for phi in Phi])
\end{minted}
\end{listing}

\subsection{Continuum solvation}
A very common approach to including solvent effects in a quantum chemistry calculation is to consider the molecule immersed in a dielectric continuum. The governing equation is the \ac{GPE}:
\begin{equation}\label{eq:gpe}
    \nabla \cdot \left[ \epsilon(\mathbf{r}) \nabla V(\mathbf{r}) \right] = - 4 \pi \rho(\mathbf{r})
\end{equation} 
where $V$ is the electrostatic potential, $\rho$ is the charge distribution of the molecule and $\epsilon$ is the (position-dependent) permittivity of the dielectric. The practical solution of the equation depends on how $\epsilon$ is parametrized. Traditionally, a cavity boundary is defined and the permittivity is a step function of the cavity: $\epsilon_i=1$ inside of it and $\epsilon_o=\epsilon_s$ outside, where $\epsilon_s$ is the bulk permittivity of the solvent. This parametrization leads to an equivalent boundary integral equation, solved with a suitable boundary element method.\cite{Tomasi2005} The main advantage of the approach is transferring the problem from the (infinite) 3-dimensional space to the (finite) cavity surface. The main disadvantage is a less physical parametrization of the model, due to the presence of the sharp boundary. 

Using \acp{MW}, Equation~\ref{eq:gpe} can instead be solved directly, without assumptions on the functional form of $\epsilon$. The main working equation is the application of the Poisson operator $\hat{P}$ to the effective charge density $\rho_\mathrm{eff}(\mathbf{x})$ and the surface polarization $\gamma(\mathbf{x})$:
\begin{equation}
    V(\mathbf{x})  = \hat{P}\left[ \rho_\mathrm{eff}(\mathbf{x}) + \gamma(\mathbf{x}) \right], \label{eq:V_R}
\end{equation}
where the effective density $\rho_\mathrm{eff}$ and the surface polarization $\gamma(\mathbf{x})$ are defined respectively as:
\begin{align}
    \rho_{eff}(\mathbf{x}) &= \frac{\rho(\mathbf{x})}{\varepsilon(\mathbf{x})} \\
    \gamma(\mathbf{x}) &= \frac{\nabla \log \epsilon \cdot \nabla V}{4 \pi}.
\end{align}
The surface polarization depends on the potential, which implies that the equation must be solved iteratively.


The equations above are easily represented in a \ac{MW} framework and each of them can be written as a single line of code with \vampyr{}. This is shown in Listing~\ref{lst:SCRF}.
The model is extensively discussed in a previous paper\cite{Gerez2023} where a thorough study of theory and implementation was presented.
The example in Listing~\ref{lst:SCRF} shows how continuum solvation has been implemented using \vampyr{}. Here we have assumed we already have a projected \verb|permittivity| 
function and the total solute \verb|density| function together with a 
suitable \verb|mra| and a function that computes $\gamma$ called \verb|computeGamma|.

The solvation energy $E_R$ is finally computed from the reaction potential $V_R = V - \int d \mathbf{r}^\prime \frac{\rho(\mathbf{r}^\prime)}{|\mathbf{r} -\mathbf{r}^\prime|}$, and the solute charge 
density $\rho$:
\begin{equation}
    E_R = \frac{1}{2} \int d\mathbf{x}V_R(\mathbf{x})\rho(\mathbf{x}).
\end{equation}

\begin{listing}[!ht]
\caption{Python implementation of PCM in \vampyr{}. }\label{lst:SCRF}
\begin{minted}{python}
# Initialize the poisson operator
P = vp.PoissonOperator(mra=mra, prec=1.0e-5)
# Solve the standard Poisson equation
V_vac = P(4*pi*density)
# Compute the initial gamma_s
gamma_s = computeGamma(V_vac)
#  Construct effective density rho_eff
rho_eff = (4 * np.pi) * ((density * (permittivity)**(-1)) - density)
# Compute the initial reaction potential
V_R = P(rho_eff + gamma_s)
# Solve the GPE by iteration
for i in range(100):
    V = V_vac + V_R
    gamma_s = computeGamma(V, permittivity)
    V_R_np1 = P(rho_eff + gamma_s)
    dV_R_n = V_R_np1 - V_R
    update = dV_R_n.norm()
    V_R = V_R_np1.deepCopy()
    if (abs(update) <= 1.0e-5):
        break        
# compute the final energy and finish. 
E_R = (1/2)*vp.dot(V_R, rho)
\end{minted}
\end{listing}

\subsection{Implementation of 4-component Dirac Equation for one-electron}

Lately we have made use of \vampyr{} to extend our work towards the relativistic domain. As such \vampyr{} has been used for a range of proof-of-principle calculations which all stem from the Dirac equation.\cite{Tantardini2024-np, Tantardini2023-vc} It is well known that the level of complexity required to deal with the Dirac equation increases significantly: the scalar non-relativistic eigenvalue problem, dealing generally with real-valued functions, becomes a 4-component problem involving complex functions:
\begin{equation}
\begin{pmatrix}
(V + mc^2- \epsilon) & c (\mathbf{\sigma} \cdot \mathbf{p}) \\
c (\mathbf{\sigma} \cdot \mathbf{p}) & (V - \epsilon -mc^2)
\end{pmatrix}
\begin{pmatrix}
\psi^{\mathrm{L}} \\
\psi^{\mathrm{S}}
\end{pmatrix} = 0
\end{equation}
where $V$ is the external potential, $\epsilon$ is the energy eigenvalue, $c$ is the speed of light, $\mathbf{p}$ is the momentum operator and $\sigma$ is the vector collecting the three Pauli matrices. As shown by Blackledge and Babajanov\cite{Blackledge2013-jh} and later exploited by \citeauthor{Anderson2019-od}\cite{Anderson2019-od}, the Dirac equation can also be reformulated as an integral equation as:
\begin{equation}
\begin{pmatrix}
\psi^{\mathrm{L}} \\
\psi^{\mathrm{S}}
\end{pmatrix} = 
\frac{-1}{2mc^2}
\begin{pmatrix}
(\epsilon + mc^2) & c (\mathbf{\sigma} \cdot \mathbf{p}) \\
c (\mathbf{\sigma} \cdot \mathbf{p}) & (\epsilon -mc^2)
\end{pmatrix} \hat{G}^{\mu} \left[ V 
\begin{pmatrix}
\psi^{\mathrm{L}} \\
\psi^{\mathrm{S}}
\end{pmatrix}
\right]
\end{equation} 
where $\hat{G}^\mu$ is the \ac{BSH} kernel as in the non relativistic case (see Section \ref{sec:conv-operators}), and $\mu = \sqrt{\frac{m^2 c^4 - \epsilon}{mc^2}}$. 

The above integral equation can be solved iteratively as in the non-relativistic case and the algorithm is thus very similar to the non-relativistic case.
The main difference is that the infrastructure required to deal with 4-component spinors needs to be implemented.
Our design is built essentially on two Python classes: one to deal with complex functions (a single component of a spinor is a complex function) and another one to deal with the four components. Thanks to \vampyr{}, these classes are easy to implement and use, for the most part overloading \emph{dunder} operators. 

    The Listing \ref{lst:orbital4c} shows a portion of the class implementation for the 4-component \texttt{Spinor}. The objects are made callable and work as \numpy{} arrays. Each component is itself a complex function object (defined in a separate class). Some \emph{dunder} methods are shown and the methods to perform the operation $c (\alpha \cdot \mathbf{p}) \psi$ are also reported.
    \begin{listing}[!ht]
    \caption{Excerpt of the 4-component spinor class.}
    \label{lst:orbital4c}
    \begin{minted}{python}
    class Spinor:
        """Four components orbital."""
        mra = None
        light_speed = -1.0
        comp_dict = {'La': 0, 'Lb': 1, 'Sa': 2, 'Sb': 3}
        def __init__(self):
            self.comp_array = np.array([cf.complex_fcn(),
                                        cf.complex_fcn(),
                                        cf.complex_fcn(),
                                        cf.complex_fcn()])
        def __getitem__(self, key):
            return self.comp_array[self.comp_dict[key]]
        def __setitem__(self, key, val):
            self.comp_array[self.comp_dict[key]] = val
        def __len__(self):
            return 4
        def __add__(self, other):
            output = orbital4c()
            output.comp_array = self.comp_array + other.comp_array
            return output
        def __call__(self, position):
            return [x(position) for x in self.comp_array]
        def __rmul__(self, factor):
            output = orbital4c()
            output.comp_array =  factor * self.comp_array
            return output
        def alpha(self, direction, prec):
            out_orb = orbital4c()
            alpha_order = np.array([[3, 2, 1, 0],
                                    [3, 2, 1, 0],
                                    [2, 3, 0, 1]])
            alpha_coeff = np.array([[ 1,  1,   1,  1],
                                    [-1j, 1j, -1j, 1j],
                                    [ 1, -1,   1, -1]])
            for idx in range(4):
                coeff = alpha_coeff[direction][idx]
                comp = alpha_order[direction][idx]
                out_orb.comp_array[idx] = coeff * self.comp_array[comp]
                out_orb.comp_array[idx].crop(prec)
            return out_orb
        def alpha_p(self, prec, der = "ABGV"):
            out_orb = orbital4c()
            orb_grad = self.gradient(der)
            apx = orb_grad[0].alpha(0, prec)
            apy = orb_grad[1].alpha(1, prec)
            apz = orb_grad[2].alpha(2, prec)
            result = -1j * (apx + apy + apz)
            result.cropLargeSmall(prec)
            return result
\end{minted}
\end{listing}

With the \texttt{Spinor} class implementation at hand, the iterative scheme to converge the Hydrogen atom is easily implemented:

\begin{listing}[!ht]
\caption{Iterative solution of the Dirac equation.}
\label{lst:integralDirac}
\begin{minted}{python}
while orbital_error > prec:
    hd_psi = orb.apply_dirac_hamiltonian(spinor_H, prec, der = default_der)
    v_psi = orb.apply_potential(-1.0, V_tree, spinor_H, prec) 
    add_psi = hd_psi + v_psi
    energy = (spinor_H.dot(add_psi)).real
    mu = orb.calc_dirac_mu(energy, light_speed)
    tmp = orb.apply_helmholtz(v_psi, mu, prec)
    tmp.crop(prec/10)
    new_orbital = orb.apply_dirac_hamiltonian(tmp, prec, energy, der = default_der) 
    new_orbital.crop(prec/10)
    new_orbital.normalize()
    delta_psi = new_orbital - spinor_H
    orbital_error = (delta_psi.dot(delta_psi)).real
    print('Error',orbital_error, imag, flush = True)
    spinor_H = new_orbital
\end{minted}
\end{listing}
First the energy is computed in order to obtain the parameter $\mu$, then comes the convolution of $V\psi$ with the Helmholtz kernel $\hat{G}^\mu$, and finally the Dirac Hamiltonian plus the energy $h_D + \epsilon$ is applied to the convolution. The iteration is repeated until the norm of the spinor update is below the requested threshold. At the end of the iteration the energy must be computed once more (not reported in the listing).

\subsection{Time-dependent Schr\"odinger equation}

The \acp{MW} framework can successfully be employed for real-time simulations of a wavepacket by directly integrating the time-dependent Schrödinger equation:
\begin{equation}
\label{general_Schrodinger}
    i \partial_t \Psi
    =
    \left( \hat{T} + \hat{V} \right) \Psi
    ,
\end{equation}
where the potential $\hat{V}$ may be time-dependent. For simplicity we consider here a time-independent potential and a one-dimensional problem:
\(
    V = V(x)
\)
and
\(
    \Psi = \Psi(x, t)
\)
with
\(
    x, t \in \mathbb R
    ,
\)
and we assume that the initial conditions are given as $\Psi(x, 0) = \Psi_0(x)$.
The standard numerical treatment of \eqref{general_Schrodinger}
is to choose a small time step $t > 0$ and construct the time evolution operator
on the interval $[0, t]$. By applying it iteratively, the solution at any finite time can in principle be obtained.
The wave propagation
can be expressed as
\begin{equation}
\label{time_propagation}
    \Psi(t)
    =
    \exp
    \left(
        -i t \left( \hat{T} + \hat{V} \right)
    \right)
    \Psi_0
    ,
\end{equation}
where we used the fact that the potential is time-independent.
We proceed by splitting the propagator in the kinetic $\exp \left(-i t \hat{T} \right)$
and potential $\exp \left(-i t \hat{V} \right)$ parts.
The former is a multiplicative operator in momentum space,
whereas the latter is a multiplicative operator in real space.
This separation is not exact, because $\hat{T}$ and $\hat{V}$ do not commute. The resulting operator has an error of order $\mathcal O \left( t^2 \right)$:
\begin{equation}
        \exp
    \left(
        -i t \left( \hat{T} + \hat{V} \right)
    \right)
    =    
    \exp
    \left(
        -i t \hat{T}
    \right)
    \exp
    \left(
        -i t \hat{V}
    \right)
    +
    \mathcal O \left( t^2 \right).
\end{equation}
This simple splitting is too rough for practical applications, therefore 
we make use of the following fourth order scheme \cite{Chin_Chen2001}

\begin{equation}
\label{Chin_Chen_A}
    e^{ At + Bt }
    =
    \exp \left(  \frac t6  B \right)
    \exp \left(  \frac t2  A \right)
    \exp \left(  \frac {2t}3  \widetilde B \right)
    \exp \left(  \frac t2  A \right)
    \exp \left(  \frac t6  B \right)
    +
    \mathcal O \left( t^5 \right)
    ,
\end{equation}
where
\begin{equation}
\label{tildeB}
    \widetilde B
    =
    B
    +
    \frac{t^2}{48}
    [ B, [A, B]]
    .
\end{equation}
With $A = -i \hat T$ and $B = -i \hat V$, $\widetilde{B}$ becomes a multiplicative operator containing the potential gradient $\partial_x V(x)$.
Remarkably, this high order scheme requires only two applications
of the free-particle semigroup operator
\(
    \exp \left(  -it  \hat T / 2 \right)
\)
per time step.
For an example of the implementation see Listing \ref{lst:chinchen}.

\begin{listing}
\caption{ChinChen implementation.}
\label{lst:chinchen}
\begin{minted}{python}
class ChinChenA(object):
    def __init__(self, expA, expB, exp_tildeB):
        self.expA = expA
        self.expB = expB
        self.exp_tildeB = exp_tildeB
    def __call__(self, u):
        u = self.expB(u)        # 1/6*dt
        u = self.expA(u)        # 1/2*dt
        u = self.exp_tildeB(u)  # 2/3*dt
        u = self.expA(u)        # 1/2*dt
        u = self.expB(u)        # 1/6*dt
        return u
\end{minted}
\end{listing}

Complex algebra is not supported natively in \vampyr{}, therefore we represent complex exponential operators as follows
\begin{equation}
    \exp \left( -i \widehat H t \right)
    =
    \begin{pmatrix}
        \cos \widehat H t
        &
        \sin \widehat H t
        \\
        - \sin \widehat H t
        &
        \cos \widehat H t
    \end{pmatrix}
    , \quad
    \mbox{operating on vector-functions }
    \Psi(t)
    =
    \begin{pmatrix}
        u(t)
        \\
        v(t)
    \end{pmatrix}
\end{equation}
where self-adjoint $\widehat H$ stands for either the kinetic energy $-\partial_x^2$, or the potential $V(x)$ or the full Hamiltonian $-\partial_x^2 + V(x)$. The implementation is shown in Listing \ref{lst:unitary_exponent_group}.
\begin{listing}
\caption{Unitary exponent group}
\label{lst:unitary_exponent_group}
\begin{minted}{python}
class UnitaryExponentGroup(object):
    def __init__(self, real, imag):
        self.real = real
        self.imag = imag
    def __call__(self, psi):
        u = psi[0]
        v = psi[1]
        res0 = self.real(u) - self.imag(v)
        res1 = self.imag(u) + self.real(v)
        return np.array([ res0, res1 ])
\end{minted}
\end{listing}
%
As a worked-out example, we show a simulation of the time evolution of a Gaussian wave packet $\Psi(x,t)$ in harmonic potential $V(x)$:
\begin{equation*}
    \Psi_0(x) = \Psi(x, t=0) =
    \left(
        \frac 1{2 \pi \sigma^2}
    \right)^{1/4}
    \exp
    \left(
        - \frac {(x - x_0)^2}{4 \sigma^2}
    \right), \qquad V(x) = V_0 (x - x_1)^2
    .
\end{equation*}

It is well known that the density
\(
    \left| \Psi(t) \right|^2
\)
oscillates in the harmonic potential with the period
\(
    \tau = \pi / \sqrt{V_0}
    .
\)
More precisely,
\(
    \Psi \left( \tau \right) = - \Psi_0
    .
\)
This can immediately be seen taking into account that
the eigenvalues for the Hamiltonian are
\(
    \sqrt{V_0} ( 2n + 1)
    .
\)
We take $x_0 = 0.3$, $\sigma = 0.04$ and
$x_1 = 0.5$, $V_0 = 25000$.
The parameters are chosen in such a way that the solution
stays localized mainly on the $[0, 1]$ interval.
Note that the
\(
    \widetilde B
\)
operator
simplifies to
\[
    \widetilde B
    =
    -i \widetilde V
    =
    -i V
    +
    \frac{it^2}{24}
    ( \partial_x V )^2
    =
    -i \widetilde V_0 (x - x_0)^2
    , \quad
    \mbox{where }
    \widetilde V_0
    =
    V_0
    -
    \frac{(tV_0)^2}6
    ,
\].
%
%

\begin{figure}
    \subfigure
    [
        Density evolution in the harmonic potential.
        \label{fig:time_solution}
    ]
    {
        \includegraphics[width=0.45\textwidth]{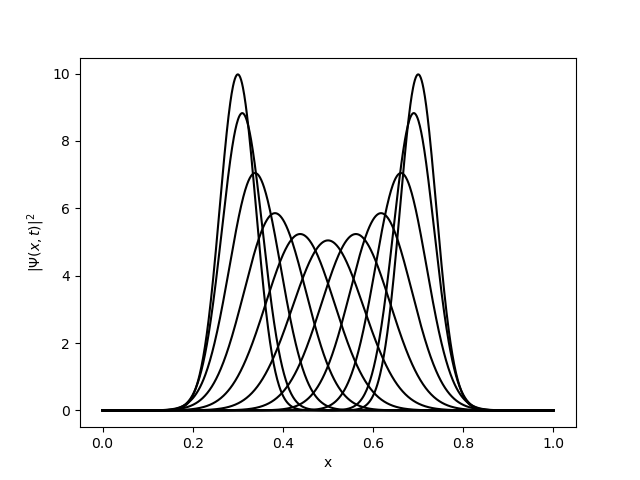}
    }
    \hfill
    \subfigure
    [
        The difference between the numerical and exact solution.
        \label{fig:time_error}
    ]
    {
        \includegraphics[width=0.45\textwidth]{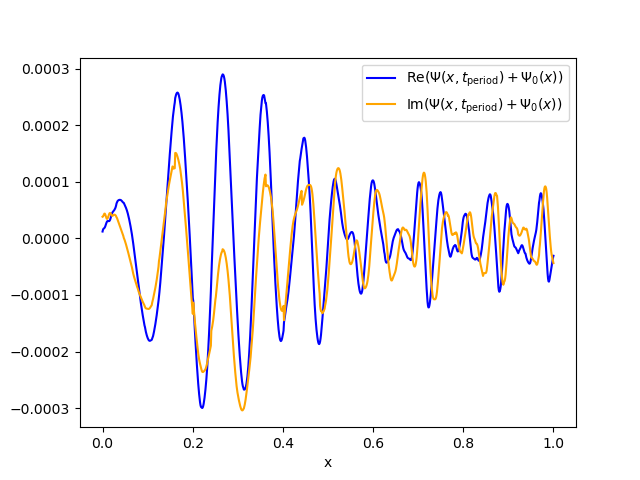}
    }
    \caption{Simulation of \eqref{general_Schrodinger} in \vampyr{}.}
    \label{fig:time_evolution}
\end{figure}

In Listing \ref{lst:preparation_time_evolution} we define all the necessary operators
and initialize Scheme \eqref{Chin_Chen_A}.
The numerical integration of the time evolution is performed in Listing \ref{lst:time_evolution_simulation}.
Figure \ref{fig:time_evolution} shows the result of the simulation:
the oscillation movement of the density
\(
    \left| \Psi(t) \right|^2
\)
and the difference between the numerical and exact solution at the time moment
\(
    t = \tau    .
\)

\begin{listing}[!ht]
\caption{Preparation to implementation of time evolution in \vampyr{}. }
\label{lst:preparation_time_evolution}
\begin{minted}{python}
# Define parameters, final time moment and time step
x0 = 0.3
sigma = 0.04
x1 = 0.5
V0 = 25000
N = 20
t_period = np.pi / np.sqrt(V0)
time_step = t_period / N
# Set the precision and make the MRA
precision = 1.0e-5
finest_scale = 9
mra = vp1.MultiResolutionAnalysis(vp1.BoundingBox(0), LegendreBasis(5))
# Make the scaling projector
P = vp1.ScalingProjector(mra, prec = precision)
# Define the harmonic potential with its scheme modification
def V(x):
    return V0 * (x[0] - x1)**2
def tilde_V(x):
    A = V0 - ( time_step * V0 )**2 / 6.0
    return A * (x[0] - x1)**2
# Define the iteration procedure
iteratorA = ChinChenA(
    expA = create_unitary_kinetic_operator(mra, precision, time_step / 2, finest_scale),
    expB = create_unitary_potential_operator(P, V, time_step / 6),
    exp_tildeB = create_unitary_potential_operator(P, tilde_V, 2 * time_step / 3)
)
\end{minted}
\end{listing}

\begin{listing}[!ht]
\caption{Time evolution simulation in \vampyr{}. }
\label{lst:time_evolution_simulation}
\begin{minted}{python}
# Define the initial wave function
psi0 = vp1.GaussFunc(
    beta = 1.0 / (4 * sigma**2), alpha = (2 * np.pi * sigma**2)**(-1/4), position = [x0]
)
psi0 = np.array([ P(psi0), vp1.FunctionTree(mra).setZero() ])
# Solve the initial value problem
psiA = psi0
for n in range(N):
    psiA = iteratorA(psiA)
# Find error at t = period
per_errorA = psiA + psi0
print( f"L2-norm of real part error:      {per_errorA[0].norm()}" )
print( f"L2-norm of imaginary part error: {per_errorA[1].norm()}" )
\end{minted}
\end{listing}

\section{Interoperability with other packages}
\label{subsec:interop}

One of the main benefits of introducing a Python interface is the seamless 
interoperability it offers with a number of other packages. This not only simplifies the 
usage of the package but also enhances its capabilities by allowing the integration of features provided by other 
packages. In this section, we will explore some potential applications of this interoperability, showcasing how \vampyr{} 
can interact with other packages to perform tasks that may otherwise be computationally demanding or 
require a more complex implementation.


As an example, the \ac{SCF} solver implemented in \vampyr{}, though fully general, encounters practical 
limitations with more complex molecular systems. One primary challenge is the selection of an appropriate initial guess 
for the orbitals; a poor choice can severely impact the convergence rate and even result in a failure to converge. The first step beyond the \ac{HF} 
approximation might be to incorporate exchange and correlation density functionals, which, while beneficial, can be 
tedious to implement.

Here is where interoperability steps in. We can leverage the capabilities of different Python packages to address these issues. We will walk through examples demonstrating how \vampyr{} can:

\begin{itemize}
\item Use \veloxchem{}\cite{Rinkevicius2020-hj} to generate an initial guess for a SCF, speeding up the process and increasing the chance of successful convergence.
\item Use \vampyr{} and \veloxchem{} to compute a potential energy surface grid, effectively utilizing computational resources.
\item Perform stability analysis on Density Functional Theory (DFT) functionals from Libxc,\cite{Lehtola2018-uc}.
\end{itemize}


\subsection{Generating an Initial Guess with \veloxchem}


Here, we illustrate how to generate an initial guess using \veloxchem{} that can be imported into \vampyr{}. Reconstructing
the \acp{MO} (or electron density) from the \ac{AO} basis set using the \ac{MO} (or density) matrix is a rather complex task,
especially if the basis contains high angular momentum functions. We thus want to make use of \veloxchem{} own internal evaluator
for these objects, and wrap a simple function around it which can be projected onto the \ac{MW} basis using a \texttt{ScalingProjector}.
In principle, any $\mathbb{R}^3 \rightarrow \mathbb{R}$ function can be projected in this way, so we just have to define a function
that takes as argument a point in real-space, runs it through \veloxchem{} internal AO evaluator code, and returns the function value
of a given \ac{MO} at that point.

First, we set up and run the SCF calculation with \veloxchem{}:
\begin{minted}{python}
# Read molecule and basis
molecule = vlx.Molecule.read_str(mol_str)
basis = vlx.MolecularBasis.read(molecule, "PCSEG-1")
# Make SCF and run driver
scf_drv = vlx.ScfRestrictedDriver()
scf_results = scf_drv.compute(molecule, basis)
# Make a visualization driver that can be used to evaluate orbitals.
vis_drv = vlx.VisualizationDriver()
# Get the MO coefficients
mol_orbs = scf_drv.mol_orbs
\end{minted}
We can define a function that returns the value of the MO at a given point in space. This function will be used for 
projecting onto an MRA in \vampyr{}:
\begin{minted}{python}
# Define a function to get the MOs, this can be projected onto an MRA by vampyr
def mo_i(mol_orbs, r):
    R = np.array([r]) 
    return vis_drv.get_mo(R, molecule, basis, mol_orbs, i, "alpha")[0]
\end{minted}
Finally, we use a projector from \vampyr{} to project the MOs from \veloxchem{} onto an MRA, generating a list of
\verb|FunctionTree|s for 
the initial guess:
\begin{minted}{python}
# This can now be imported into vampyr as:
P_eps = vp.ScalingProjector(mra, prec)
Phi_0 = [P_eps(mo_i) for i in range(nr_orbs)]
\end{minted}
In the last step, we loop over the desired number of orbitals, project each one onto the MRA, and store the result in 
\verb|Phi_0|. This list of \verb|FunctionTree|s represents an approximation to the molecular orbitals of the 
system, and serves as an 
excellent initial guess for the SCF procedure in \vampyr{}. 


\subsection{Calculate a potential energy surface with \veloxchem{} and \vampyr{}}

A \ac{PES} can be computed as a series of single point energy calculations while varying one (or more)\footnote{In this example we do it with
one variable. But current \vampyr{} can support it up to 3 variables.} bond distance(s)
in a molecule. We will here use \vampyr{}'s adaptive function projector as a driver for computing the \ac{PES} of the
Hydrogen molecule using \veloxchem{}'s single-point Hartree--Fock evaluator. We define a Python function, \verb|pes(r)|,
that computes the energy for two Hydrogen atoms given the bond distance as input:

\begin{minted}{python}
def pes(r):
    mol_str = f""" 
    H       0.0  0.0000   -0.2
    H       0.0  0.0000   {r[0]}
    """
    molecule = vlx.Molecule.read_str(mol_str)
    basis = vlx.MolecularBasis.read(molecule, "PCSEG-1")

    scf_drv = vlx.ScfUnrestrictedDriver()
    scf_results = scf_drv.compute(molecule, basis)
    return scf_drv.scf_energy
\end{minted}
The function above computes the energy of a Hydrogen molecule as a function of the bond distance (the 0.2 offset is to avoid the singularity at $r=0$), by performing an SCF calculation at the given geometry.
Being a $\mathbb{R} \rightarrow \mathbb{R}$ mapping, it can be projected on a 1D \ac{MRA} using \vampyr{}.

\begin{minted}{python}
mra = vp.MultiResolutionAnalysis(box=[0, 10], order=1)
P_eps = vp.ScalingProjector(mra, prec=1.0e-3)
pes_tree = P_eps(pes)
\end{minted}

\begin{figure}
    \centering
    \includegraphics[width=\textwidth]{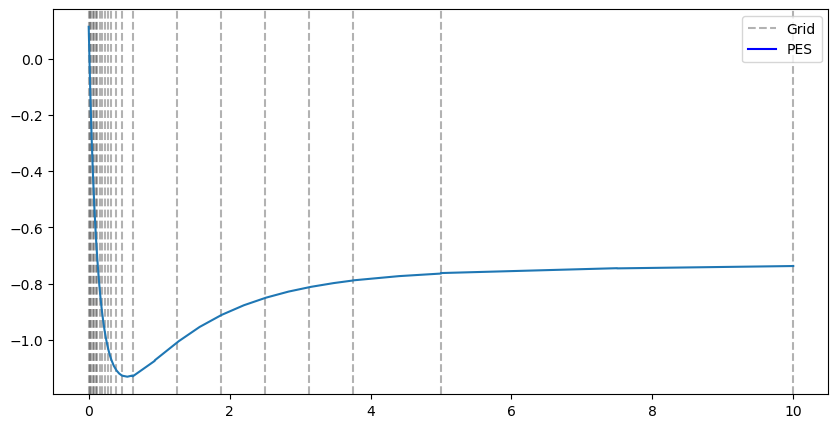}
    \caption{Potential energy surface calculated using \veloxchem{} and \vampyr{}. The grid lines represent the adaptive MRA grid, from which the PES is interpolated.}
    \label{fig:pes}
\end{figure}

The adaptive projector will automatically sample the \verb|pes(r)| function in appropriate points in order to produce a
smooth surface. The potential energy surface obtained in this manner can be visualized, as shown in Figure \ref{fig:pes}.

The vertical lines in Figure~\ref{fig:pes} represent the adaptive MRA grid, and the potential energy surface depicted is interpolated from this grid rather than computed directly at each point. The adaptive algorithm focuses computational resources where needed to achieve the desired precision. Further efficiency gains could be achieved realizing that high precision is only required for region of low potential energy. This has not been exploited in the above example.

\subsection{Numerical Stability Analysis using \vampyr{} and \pylibxc{}}

Inspired by the paper by \citeauthor{lehtola2022many},\cite{lehtola2022many} we investigate the numerical stability of density 
functional approximations (DFAs) lda-c-vwn and lda-c-gk72, with \vampyr{} and \pylibxc{}. The choice of these DFAs is due 
to their contrasting numerical stability, making them interesting subjects for our analysis. The lda-c-gk72, 
developed by Gordon and Kim in 1972, was known for its problematic convergence while the lda-c-vwn (Vosko, Wilk, and 
Nusair functional) was noted for its numerical stability. Our objective is to verify these claims, illustrating the potential of \vampyr{} as numerical analysis tool.

Firstly, we initialize a Neon atom and calculate the self-consistent field (SCF) using \veloxchem{}:
\begin{minted}{python}
def init_molecule_and_scf(mol_str, basis_set="PCSEG-1"):
    molecule = vlx.Molecule.read_str(mol_str)
    basis = vlx.MolecularBasis.read(molecule, basis_set)
    scf_drv = vlx.ScfRestrictedDriver()
    scf_results = scf_drv.compute(molecule, basis)
    return molecule, basis, scf_drv.density
\end{minted}
Next, we calculate the density and project it onto a Multi-Resolution Analysis (MRA) grid:
\begin{minted}{python}
def calc_density(r, molecule, basis, mol_orbs, vis_drv):
    R = np.array([r])
    rho_a = vis_drv.get_density(R, molecule, basis, mol_orbs, "alpha")[0] +
    rho_b = vis_drv.get_density(R, molecule, basis, mol_orbs, "beta")[0]
    return rho_a + rho_b
    
rho = P_eps(lambda r: calc_density(r, molecule, basis, mol_orbs, vis_drv))
\end{minted}
Subsequently, we use \pylibxc{} to define the chosen DFA functionals:
\begin{minted}{python}
func_c = xc.LibXCFunctional("lda_c_gk72" or "lda_c_vwn", "unpolarized")
def f_e(n):
    c = func_c.compute({"rho": n})["zk"]
    return n * (c)
F_e = vp.FunctionMap(f_e, precision)
\end{minted}
Following this, an exchange potential is generated. We iteratively refine the grid based on precision, one scale at a time, until no new 
nodes are created:
\begin{minted}{python}
v_x = vp.FunctionTree(mra)
nNodes = v_x.nNodes()
while nNodes > 0:
    vp.advanced.clear_grid(v_x)
    vp.advanced.map(-1.0, v_x, rho, f_e)
    print(f"nodes : {v_x.nNodes()} norm : {v_x.norm()}", flush=True)
    nNodes = vp.advanced.refine_grid(v_x, precision)
\end{minted}

Observing the output for each DFA, we note that the number of nodes converges to a finite value for the chosen precision in case of the \verb|lda_c_vwn| functional. For the \verb|lda_c_gk72| functional
the number of nodes grows exponentially before the requated precision is reached. Looking at the wavelet norm, we notice that we are i still fa r from a converged result. In this example we employed a rather high precision ($\epsilon = 10^{-8}$).

\begin{figure}
    \centering
    \includegraphics{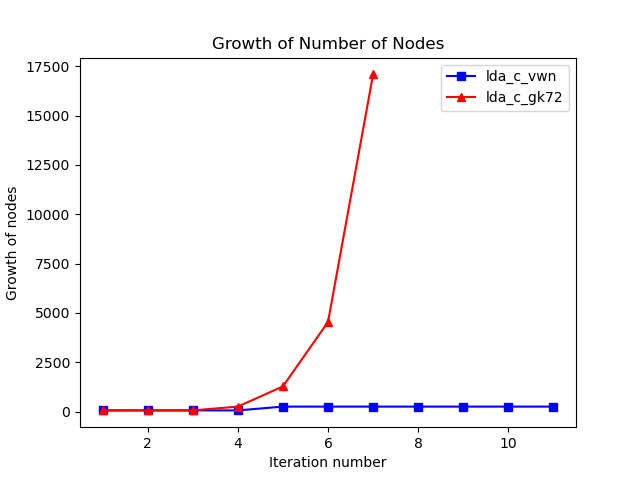}
    \caption{Comparison of node growth between the lda-c-vwn and lda-c-gk72 DFAs. The plot illustrates the sharp contrast in the convergence behavior, with lda-c-vwn showing a stable, linear growth, and lda-c-gk72 displaying an exponential growth.}
    \label{fig:pylibxc-nodes}
\end{figure}

\section{Summary}

Multiresolution analysis is a relatively new tool in quantum chemistry, compared to traditional methods based on atomic orbitals. Its main advantages are the possibility to reach any predefined arbitrary precision (the only limits are dictated by the machine precision and the available memory), and a numerical implementation which stays close to the theoretical formalism. \vampyr{} has been designed to capitalize on these two aspects, and at the same time enable fast code development for prototype applications.

We have employed \vampyr{} to test new ideas and methods, and to create interfaces with existing codes. In the first category, we have here shown a simple SCF optimization, a continuum solvation model, the solution of the Dirac equation for one electron and the application of the time evolution operator. In the second, we illustrated how to import a starting guess for the orbitals, the plotting of a \ac{PES}, a numerical analysis of two density functionals. For all the above examples we have provided extracts of the code which show how \vampyr{} is used. The complete set of working examples illustrated in Section~\ref{sec:vampyrSCF} has been collected in an openly accessible GitHub repository\cite{vampyr_coven}, under the CC-BY-4.0 license.

\appendix

\section{The Self Consistent Field equations of Hartree--Fock and Density Functional Theory}\label{sec:HFappendix}

The starting point of most Quantum Chemistry calculations involve one Slater determinant, because it automatically accounts for the Pauli exclusion principle and allows to express quantities in a sum of orbital contributions. In particular, the energy expression for a Slater determinant $\Psi$ in atomic units is given by:
\begin{align}
    E[\Psi] = \sum_i \braket{ \psi_i | \hat{h} | \psi_i} + \frac{1}{2}
              \sum_{i,j} \braket{ \psi_i | \hat{J}_j - \hat{K}_j | \psi_j }
\end{align}
Here, $\hat{h}$ is the one-electron Hamiltonian:
\begin{align}\label{eq:one-electron-op}
    \hat{h} = \hat{T} + \hat{V}_{N} = - \frac{1}{2} \nabla^2 - \sum_I \left(\frac{Z_I}{|r - R_I|}\right),
\end{align}
with $\hat{T}$ the kinetic energy operator and $\hat{V}_N$ is the attractive nuclear-electron potential energy operator.
$\hat{J}_j$ and $\hat{K}_j$ denote the Coulomb and 
exchange interaction operators, respectively, and are defined as:
\begin{align}
    \label{eq:couloumb_oper}
    \hat{J}_j \psi_i &= \hat{P}\left[ |\psi_j|^2 \right] \psi_i \\
    \hat{K}_j \psi_i &= \hat{P}\left[ \psi_j \psi_i \right] \psi_j
    \label{eq:exchange_oper}
\end{align}

By minimizing the energy $E[\Psi]$ with respect to variations in the orbitals, under the constraint that the spatial orbitals remain orthonormal:
\begin{align}
    \braket{E}_{HF} = \min E[\Psi] \qquad \braket{ \psi_i | \psi_j } = \delta_{i,j},
\end{align}
the Hartree--Fock equations are obtained\cite{jensen2017introduction}:
\begin{align}
\label{eq:general-fock}
    \hat{F}\psi_i = \left( 
    \hat{h} + \hat{J} - \hat{K} 
    \right)\psi_i = \sum_{j} F_{ij}\psi_j
\end{align}
where $\fockOper{}$ is the Fock operator and $\fockMat_{ij} = \braket{\psi_i | \fockOper | \psi_j}$ are its matrix elements in the molecular orbital basis. The canonical \ac{HF} equations are obtained by diagonalizing the Fock operator:
\begin{align}
\label{eq:canonical-fock}
    \hat{F}\phi_i = \epsilon_i \phi_i
\end{align}

Traditional quantum chemistry methods make use of an expansion of the orbitals in a fixed set of atomic orbitals $\lbrace\chi_{\alpha}\rbrace$. Each orbital is expressed as a linear combination of atomic orbitals $\phi_i = \sum_\alpha \chi_\alpha C_{\alpha i}$, where elements $C_{\alpha i}$ representing the transformation from moelcular to atomic orbitals are collected in the matrix $\mathbf{C}$. The resulting equations, after substituting the expansion in Eq.~\ref{eq:canonical-fock}, multiplying by a second atomic orbital $\chi_\beta$ and integrating, are the so-called Roothan-Hall equations:
\begin{align}
\label{eq:roothan-hall}
    \mathbf{F}\mathbf{C} = \epsilon \mathbf{S} \mathbf{C}
\end{align}

The problem is then cast in matrix form and the eigenvalues (energies) and eigenvectors (orbital expansion coefficients) are obtained by standard linear algebra techniques. 
The immediate advantage is a representation closely related to the physics of the system (atomic orbitals), but this comes at a price: the implementation deals with the representation of such functions, their integrals, their overlaps, and seemingly simple operations such as applying an operator or multiplying two functions, become technically complicated obfuscating the physical significance. Moreover the initial scaling of the problem becomes $n^4$, where $n$ is the number of basis functions, due to the two-electron integrals, and a lot of effort is necessary to recover the original $n^2$ scaling, which is instead straightforward for a \ac{MW} implementation.

When \acp{MW} are used, Eq.\ref{eq:general-fock} is instead kept in its original form, but converted to an integral equation:

\begin{align}
\label{eq:integral-fock}
    \left(
    \hat{T} + \hat{V}_N + \hat{J} - \hat{K} 
    \right)\psi_i &= \sum_{j} F_{ij}\psi_j\\
    \left(
    \hat{T} - F_{ii}
    \right) \psi_i &= 
    -\left(
    \hat{V}_N + \hat{J} - \hat{K}
    \right)\psi_i + \sum_{j \neq i} F_{ij}\psi_j\\
    \psi_i &= 
    -2\hat{G}^{\mu_i} 
    \left[
    \left(
    \hat{V}_N + \hat{J} - \hat{K}
    \right)\psi_i - \sum_{j \neq i} F_{ij}\psi_j
    \right]
\end{align}

In the equations above, we have first split the one-electron operator $\hat{h}$ in the kinetic energy $\hat{T}$ and the potential energy $\hat{V}_N$, then rearranged the terms, including the diagonal element $F_{ii}$ of the Fock matrix, and finally applied the \ac{BSH} Green's function to obtain the final expression. There is here no need to transform the problem to a different basis. The last term, which represents the Lagrangian multipliers can be omitted if the canonical \ac{HF} equations are used. The final equation can moreover be interpreted as a preconditioned steepest descent, where the gradient is the expression in the square bracket (it can be obtained by taking the differential of the expectation value of the energy of a Slater determinant with respect to a generic orbital variation) and the preconditioner is the \ac{BSH} kernel.

\section{Time Evolution Operator}

The key component to achieve an efficient quantum
time evolution simulation is a good numerical representation
of
\(
    \exp \left( i t \partial_x^2 \right)
    .
\)
We remind that the exponent operator forms a semigroup
representing solutions of the free particle equation
\[
    i \partial_t \Psi + \partial_x^2 \Psi = 0
    .
\]
In other words, for any $\Psi_0 \in L^2(\mathbb R)$
function
\(
    \Psi
    =
    \exp \left( i t \partial_x^2 \right)
    \Psi_0
\)
solves this equation.
This is a convolution operator with the kernel
\[
    K(x, y)
    =
    \frac{ \exp( -i \pi / 4 ) }{ \sqrt{4 \pi t} }
    \exp
    \left(
        \frac{ i(x - y)^2 }{4t}
    \right)
    ,
\]
and so one anticipates that the machinery described above would work here as well.
In practice it turns out to be difficult to discretise it in a similar manner without
damping the high frequencies of Green's function \cite{Vence_Harrison_Krstic2012}.
We use a different approach based on the fact that it is a multiplication operator
in the frequency domain,
namely,
\(
    \widehat \Psi(\xi, t)
    =
    \exp \left( - i t \xi^2 \right)
    \widehat \Psi_0(\xi)
    .
\)

The detailed theory behind algorithm in use follows in a separate
upcoming publication \cite{Dinvay_Frediani2023}.
Here we present only the working formulas
encoded in \vampyr{}.
Let
\(
    \left[ \sigma_{l' - l}^{n} \right]_{j'j}(t)
\)
stay for matrix elements of the time evolution operator
\(
    P^n \exp \left( i t \partial_x^2 \right) P^n
\)
at scale $n$ with respect to the Legendre scaling basis
\(
    \varphi_{j, l}^n (x)
    .
\)
Then
\begin{equation}
\label{operator_correlation_expansion}
    \left[ \sigma_l^n \right]_{pj}
    (t)
    =
    \sum_{k = 0}^{\infty}
    C_{jp}^{2k}
    J_{2k + j + p}(l, 4^n t)
    ,
\end{equation}
where
\begin{equation}
\label{power_integral}
    J_m(l, a)
    =
    \frac
    {
        e^{ i \frac {\pi}4 (m - 1) }
    }
    {
        2 \pi ( m + 2 )!
    }
    \int_{\mathbb R}
    \exp
    \left(
        \rho l \exp \left( i \frac \pi 4 \right) - a \rho^2
    \right)
    \rho^m
    d \rho
\end{equation}
satisfying the following relation
\begin{equation}
\label{power_integral_recursion}
    J_{m+1}
    =
    \frac
    {
        il
    }
    {
        2a (m + 3)
    }
    J_m
    +
    \frac {im}{2a(m + 2)(m + 3)}
    J_{m-1}
    , \quad
    m = 0, 1, 2, \ldots,
\end{equation}
with the agreement $J_{-1} = 0$ and
\begin{equation}
\label{power_integral_0}
    J_0
    =
    \frac{ e^{ -i \frac{\pi}4 } }{ 4 \sqrt{ \pi a } }
    \exp
    \left(
        \frac{il^2}{4a}
    \right)
    .
\end{equation}
These power integrals depend on the time step parameter $t > 0$,
whereas the coefficients $C_{jp}^{2k}$ are problem-independent
and can be calculated once as
\begin{equation*}
    C_{jp}^k
    =
    \sum_{m=0}^j
    \sum_{q=0}^p
    \frac{ (-1)^{m+1} (k + 2 + j + p)! }{ (k + 2 + j + p + m + q)! }
    \left(
        A_m^j B_q^p
        +
        (-1)^{k + j + p + m + q}
        B_m^j A_q^p
    \right)
    .
\end{equation*}
The coefficients appearing here under the double sum may be found as follows
\begin{equation*}
    A_0^1 = \sqrt{3}
    , \quad
    A_1^1 = 2 \sqrt{3}
    , \quad
    B_0^1 = \sqrt{3}
    , \quad
    B_1^1 = -2 \sqrt{3}
    .
\end{equation*}
For $j \geqslant 1$ we have the following relation
\begin{equation*}
\begin{aligned}
    A_0^{j+1}
    &=
    \sqrt{ \frac{ 2j + 3 }{ 2j - 1 } }
    A_0^{j-1}
    \\
    A_1^{j+1}
    &=
    \sqrt{ \frac{ 2j + 3 }{ 2j - 1 } }
    A_1^{j-1}
    -
    2 \sqrt{ (2j + 1)(2j + 3) }
    A_0^j
    \\
    &\ldots \qquad \ldots \qquad \ldots \qquad \ldots \qquad \ldots
    \\
    A_{j-1}^{j+1}
    &=
    \sqrt{ \frac{ 2j + 3 }{ 2j - 1 } }
    A_{j-1}^{j-1}
    -
    2 \sqrt{ (2j + 1)(2j + 3) }
    A_{j-2}^j
    \\
    A_j^{j+1}
    &=
    -
    2 \sqrt{ (2j + 1)(2j + 3) }
    A_{j-1}^j
    \\
    A_{j+1}^{j+1}
    &=
    -
    2 \sqrt{ (2j + 1)(2j + 3) }
    A_j^j
\end{aligned}
\end{equation*}
and $B_m^j$ obey the same recurrence for $j \geqslant 1$.

The \vampyr{} implementation of the time evolution operator
\(
    \exp \left( i t \partial_x^2 \right)
\)
is under optimization currently,
though it is already available in \vampyr{} (see Listing \ref{lst:timeevolutionoperator}).
\begin{listing}
\caption{Time evolution operator}
\label{lst:timeevolutionoperator}
\begin{minted}{python}
def create_unitary_kinetic_operator(mra, precision, time, finest_scale):
    real = vp1.TimeEvolutionOperator(mra, precision, time, finest_scale, False)
    imag = vp1.TimeEvolutionOperator(mra, precision, time, finest_scale, True)
    return UnitaryExponentGroup(real, imag)
\end{minted}
\end{listing}

Finally, the tree structure of the potential semigroup
\(
    \exp \left(  -i t  V \right)
\)
is introduced Listing \ref{lst:unitary_potential_operator}.
\begin{listing}
\caption{Unitary potential operator}
\label{lst:unitary_potential_operator}
\begin{minted}{python}
def create_unitary_potential_operator(P, V, t):
    def real(x):
        return np.cos(V(x) * t)
    real = P(real)
    def imag(x):
        return - np.sin(V(x) * t)
    imag = P(imag)
    real = MultiplicationOperator(real)
    imag = MultiplicationOperator(imag)
    return UnitaryExponentGroup(real, imag)
\end{minted}
\end{listing}
Together with a splitting scheme, for example \eqref{Chin_Chen_A},  this completes the algorithm
description for the time evolution simulations.

\bibliography{main}

\begin{thebibliography}{58}
\bibitem [{Notea()}]{Notea}%
  \BibitemOpen
  \bibinfo {note} {It should be noted that some performance-oriented functionalities, as well as specific operators like Helmholtz and Poisson, are currently exclusive to 3-dimensional problems. The extensions to lower or higher dimensions are possible, but are currently not implemented.}\BibitemShut {Stop}%
\bibitem [{Noteb()}]{Noteb}%
  \BibitemOpen
  \bibinfo {note} {In general, one root node is sufficient, but it is possible to specify a domain containing more than one root node. This option can be useful in the multidimensional case, if one wants a domain which had different sizes in the different directions.}\BibitemShut {Stop}%
\bibitem [{Noted()}]{Noted}%
  \BibitemOpen
  \bibinfo {note} {In this example we do it with one variable. But current \protect \texttt {VAMPyR}{} can support it up to 3 variables.}\BibitemShut {Stop}%
\end{thebibliography}

\end{document}